\documentclass[a4,12pt]{article}
\usepackage{type1cm,amsmath,amssymb,color,graphics,amscd,amsfonts,epsf,indentfirst}
\usepackage{epsfig}
\usepackage{subfigure}
\usepackage[dvipdfm,hypertex]{hyperref}
\newcommand{\Slash}[1]{\ooalign{\hfil/\hfil\crcr$#1$}}

\allowdisplaybreaks[4]

\makeatletter
\@addtoreset{equation}{section}

\makeatletter
\renewcommand\section{\@startsection {section}{1}{\z@}%
                                   {-3.5ex \@plus -1ex \@minus -.2ex}
                                   {2.3ex \@plus.2ex}%
                                   {\normalfont\large\bfseries}}
\renewcommand\subsection{\@startsection{subsection}{2}{\z@}%
                                     {-3.25ex\@plus -1ex \@minus -.2ex}%
                                     {1.5ex \@plus .2ex}%
                                    {\normalfont\bfseries}}

\def\no{\nonumber \\}
\def\btab{\begin{table}[h] \begin{center} \begin{tabular}{l lp{3in}}}
      \def\etab{\end{tabular} \end{center} \end{table}}
\def\btabm{\begin{center} \begin{tabular}}
    \def\etabm{\end{tabular} \end{center}}

\def\ie{{\it i.e.}}

\def\a{{\alpha}}
\def\al{\alpha'}
\def\b{{\beta}}

\def\G{{\Gamma}}
\def\d{\delta}
\def\D{{\Delta}}
\def\ep{{\epsilon}}

\def\m{{\mu}}
\def\n{{\nu}}

\def\t{{\theta}}

\def\Om{\Omega}

\def\f#1#2{{\frac{#1}{#2}}}

\def\s{\sqrt}

\def\f {\frac}
\def\ti{\tilde}

\def\p{\partial}
\def\we{\wedge}

\def\Dslash{D \llap{/\,}}

\def\CD{{\cal D}}

\def\CL{{\cal L}}
\def\CR{{\cal R}}

\def\CN{{\cal N}}

\def\BZ{\mathbb{Z}}

\begin{document}

\begin{titlepage}
  \thispagestyle{empty}
  
  \begin{flushright} 
    KUNS-2134\\
    
  \end{flushright} 
  
  \vspace{2cm}
  
  \begin{center}
    \font\titlerm=cmr10 scaled\magstep4
    \font\titlei=cmmi10 scaled\magstep4
    \font\titleis=cmmi7 scaled\magstep4
     \centerline{\titlerm
       Phase Transitions of Charged Kerr-AdS Black Holes}
    
    \vspace{0.4cm}
    \centerline{\titlerm 
      from Large-{\LARGE $N$} Gauge Theories
    }
    
    \vspace{2.2cm}
    \noindent{{
        Keiju Murata\footnote{e-mail:murata@tap.scphys.kyoto-u.ac.jp},
        Tatsuma Nishioka\footnote{e-mail:nishioka@gauge.scphys.kyoto-u.ac.jp},
        Norihiro Tanahashi\footnote{e-mail:tanahashi@tap.scphys.kyoto-u.ac.jp}
        and Hikaru Yumisaki\footnote{e-mail:yumisak@scphys.kyoto-u.ac.jp}
      }}\\
    \vspace{0.8cm}
    
   {\it Department of Physics, Kyoto University, Kyoto 606-8502, Japan} 
    
   \vspace{1cm}
   {\large \today}
  \end{center}

  \vskip 3em

  \begin{abstract}
    We study $\CN =4$ super Yang-Mills theories on a three-sphere 
    with two types of chemical potential.
    One is associated with the R-symmetry and the other with the
    rotational symmetry of $S^3$ ($SO(4)$ symmetry). 
    These correspond to charged Kerr-AdS black holes via AdS/CFT.
    The exact partition functions at zero coupling are computed and the
    thermodynamical properties are studied.
    We find a nontrivial gap between the confinement/deconfinement transition
    line and the boundary of the phase diagram 
    when we include more than four chemical potentials.
    In dual gravity, we find such a gap in the phase diagram by studying the
    thermodynamics of the charged Kerr-AdS black hole.
    This shows that the qualitative phase structures agree between
    both theories.
    We also find that the ratio between the thermodynamical quantities is
    close to well-known factor of 3/4 even at low temperatures.
  \end{abstract}

\end{titlepage}

 \noindent\rule\textwidth{.1pt}
 \vskip 2em \@plus 3ex \@minus 3ex
 \tableofcontents
 \vskip 2em \@plus 3ex \@minus 3ex
 \noindent\rule\textwidth{.1pt}
 \vskip 2em \@plus 3ex \@minus 3ex


\section{Introduction and summary}

The AdS/CFT correspondence has played a central role for about ten years in
the study of the strongly coupled region of $\cal N$ $=4$ super 
Yang-Mills theory with $SU(N)$ gauge group because it is
simply described by type IIB supergravity on $AdS_5 \times S^5$ 
\cite{Ma,GKP,Wi} (see \cite{AGMOO} for a review). 
It is well-known that the thermodynamical quantities in free gauge theory
agree with those in dual gravity up to a factor of $3/4$ \cite{GKP2,Wi2}.
This factor does not change if we consider other SYM theories such as
the one with R-symmetry chemical potentials (dual to R-charged black
holes)\cite{Gu,CG,HaOb} or the others with $SO(4)$ symmetry chemical 
potentials associated with the angular momenta of fields on a 
three-sphere (dual to Kerr-AdS black holes) \cite{BP,HHT}.
Also it has been shown quantitatively \cite{KoRi,NiTa} 
that this discrepancy is always nearly $3/4$ for infinitely many $\cal
N$ $=1$ SCFTs, which can be constructed systematically
\cite{HV,FHMSVW} from dual $AdS_5 \times Y_5$ geometries, 
where $Y_5$ is a toric Sasaki-Einstein manifold \cite{GMSW,CLPP}.
This agreement suggests that the free approximation of gauge
theory captures the significant properties in the strongly coupled
theory if the AdS/CFT correspondence holds. 

In the AdS space, there is a phase transition between the thermal AdS space
and the AdS-Schwarzschild black hole, the so-called Hawking-Page transition
\cite{HP}.  It has been pointed out in \cite{Wi2} 
that this corresponds to the
confinement/deconfinement transition in the strongly coupled gauge theory. 
Although naively there seems to be no phase transition in the gauge theory
defined on a compact space $S^3$ since we have only a finite degree
of freedom, we know of such a example in the large-$N$ limit:
the Gross-Witten-Wadia transition \cite{GW,Wa}.
The infinite degree of freedom in the large-$N$ limit causes the phase 
transition even in a finite-volume system.
It was shown in \cite{Su,AMMPV} that there exists 
such a phase transition even at zero coupling by studying large-$N$ 
gauge theories with constituent states in the adjoint
representation of the gauge group. 
In the absence of any interaction, but with a singlet (Gauss' law) constraint 
considered, the gauge theory becomes an exactly solvable unitary matrix
model of the Polyakov loop. 
This model exhibits a phase transition such that the expectation value of 
the Polyakov loop is zero below some critical temperature and 
becomes nonzero above it. 
Also the free energy scales as $O(1)$ in the low-temperature phase and
$O(N^2)$ in the high-temperature phase. 
This is precisely the confinement/deconfinement phase 
transition. 
This model has been extensively studied in \cite{Liu,AGLW,ABMW,Hi,ABW,DG} 
including weak coupling, the finite-$N$ effect and the orbifold.

Similar analysis has been performed regarding the presence of the
global R-symmetry chemical potentials in \cite{YY,BW,HO,HO2}, 
and a weak coupling region with near-critical chemical 
potentials has recently been investigated \cite{HKNW}.
The resulting phase diagram of the zero-coupling limit is very similar to 
that of the gravitational solutions of five-dimensional $\cal N$ $=2$ 
gauged supergravity.

In this paper, we study $\cal N$ $=4$ SYM theory on a three-sphere 
with general (R- and $SO(4)$ symmetry) chemical potentials following 
the method in \cite{Su,AMMPV}.\footnote{
Similar setup has been
  studied in the decoupling limit in \cite{HKO}.}
The dual theory to this gauge theory 
is the five-dimensional maximal $SO(6)$-gauged $\mathcal{N}=8$ supergravity.
The $SO(6)$ gauge symmetry, which originates from the $S^5$ compactification
of ten-dimensional type IIB supergravity, incorporates $U(1)^3$ symmetry.
This symmetry corresponds to the R-symmetry of SYM theory on $S^3$.
Thus, we should compare the solutions in this five-dimensional $U(1)^3$-gauged 
$\mathcal{N}=2$ supergravity theory with the fields 
in SYM theory with R-charge chemical potentials.
Despite considerable effort devoted toward finding
the exact black hole solution within this
theory~\cite{CvLuP,CvLP,CGLP,KuLu,CCLP3,CCLP,CLP,KLR2,CCLP2,MP}, 
the most general solution with three charges and two angular momenta
has not been yet found.
However, we can construct the most general dual gauge theory with 
three R-symmetry chemical potentials and two $SO(4)$ chemical potentials.
Therefore, we may expect that the various properties of the 
undiscovered black hole solution can be induced from the analysis of
dual gauge theory.
In this paper we focus on a solution of the five-dimensional
charged Kerr-AdS black hole constructed in \cite{CvLP,CGLP}, 
which has two equal angular momenta and three independent charges.

Setting the R-symmetry chemical potential to zero, we obtain 
$\cal N$ $=4$ SYM with $SO(4)$ chemical potentials. 
This has already been considered in \cite{HR}, and 
the AdS/CFT correspondence about the Kerr-AdS black hole was investigated.
However, the constraint of Gauss' law was not taken into account, which plays
a crucial role in gauge theory on a compact space.
Therefore the analysis is only valid at the high-temperature limit where
the compact space can be approximated to a flat space.
It is necessary to maintain Gauss' law even at zero coupling
 to obtain valid results in the limit of the interacting theory
as pointed out in \cite{YY}.
We will consider the Gauss' law constraint in the analysis of gauge theory,
and show that the confinement/deconfinement transition occurs at zero coupling.
This transition  was not observed in the analysis by \cite{HR}.

\begin{figure}[htbp]
  \centering
  \subfigure[Yang-Mills (zero coupling)]{
    \includegraphics[scale=0.4]{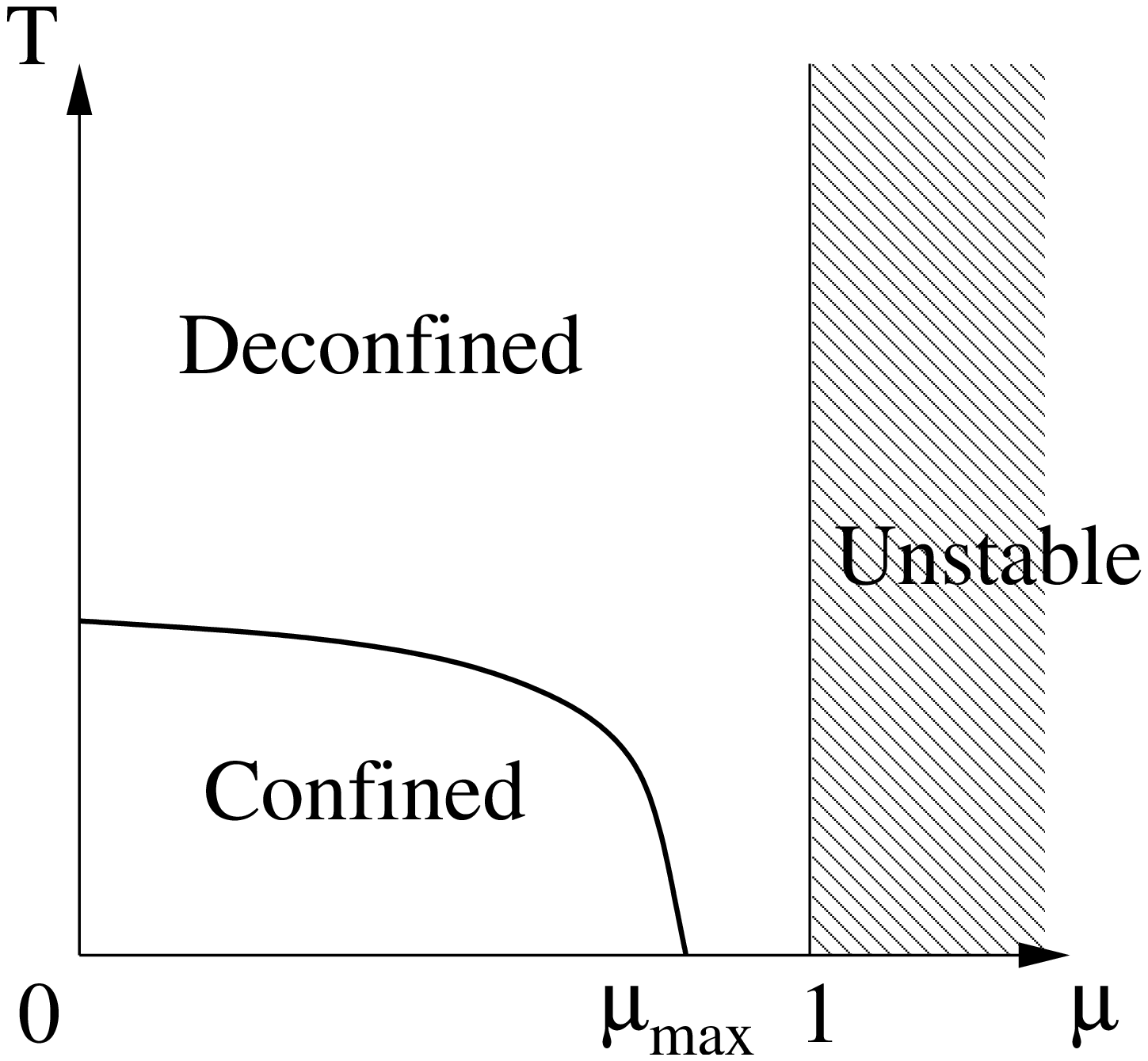}\label{fig:CR-phase}
  }
  \hspace{0.8cm}
  \centering
  \subfigure[Gravity (strong coupling)]{
    \includegraphics[scale=0.4]{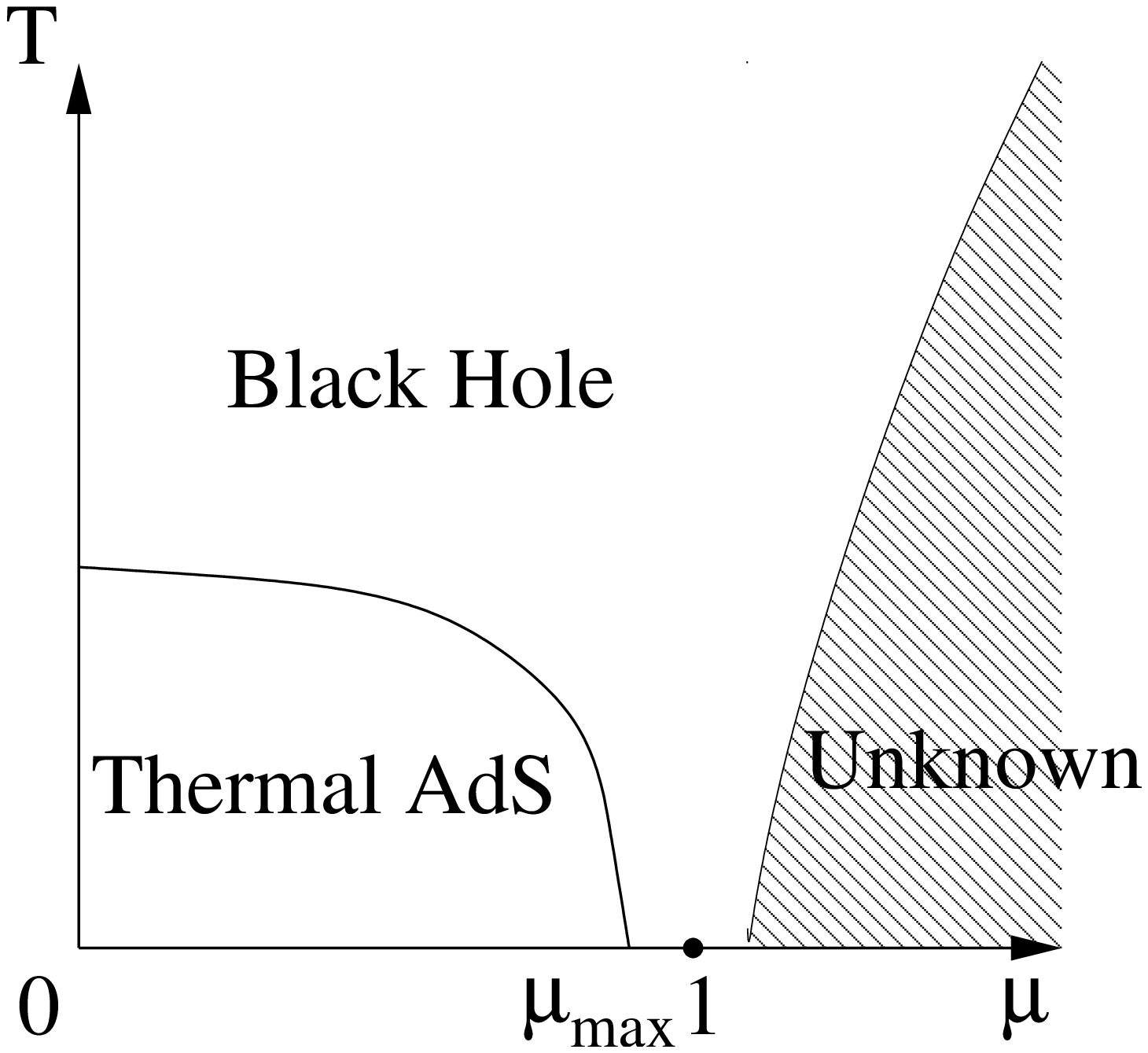}\label{fig:CRBH-phase}
  }
  \caption{Phase diagrams of $\CN =4$ large-$N$ SYM theory
    (a) and the charged Kerr-AdS black hole (b).
    We take $(\Omega_1,\Omega_2,\mu_1,\mu_2,\mu_3)=(0.9,0.9,\mu,\mu,0)$,
    so that we can see the typical features of general
    phase diagrams.
  }
  \label{fig:CR-Com}
\end{figure}

The organization of this paper is as follows. In section 2, 
we study $\CN =4$ SYM theory on a three-sphere
with chemical potentials associated with the R-symmetry of $\CN =4$
supersymmetry and the $SO(4)$ symmetry of $S^3$.
The $SU(4)$ R-symmetry has a $U(1)^3$ Cartan subalgebra; thus, we 
can introduce the three chemical potentials $(\m_1,\m_2,\m_3)$ discussed
in \cite{YY}, while the $SO(4)$ symmetry has a $U(1)^2$ Cartan
subalgebra, so we have two associated chemical potentials $(\Om_1,\Om_2)$. 
Then we construct a partition function for free $\cal N$ $=4$ SYM with 
these chemical potentials following \cite{Su,AMMPV},
and we determine the phase diagram in the phase
space $(\Om_1,\Om_2,\m_1,\m_2,\m_3)$ (Fig.\ \ref{fig:CR-phase}).
In this diagram, we have found the maximal chemical potential
$\m_{\text{max}}$ as a function of the other chemical potentials,
below which the entire body of the transition line is enclosed 
in the phase diagram.
The appearance of $\mu_{\text{max}}$ is related to the 
divergence of the fermion partition function, but
the theory is still valid because of Pauli exclusion principle.
We also determine upper bounds for the chemical potentials
above which some field becomes tachyonic.
We call this boundary line
the unitarity line since the unitarity of the theory 
breaks down above it.
We show that a gap appears between the confinement/deconfinement
transition line and the unitarity line when there are more than four
chemical potentials in the gauge theory.
This is a new phenomenon discovered in this paper.
The theories with only R-symmetry or $SO(4)$ symmetry are included 
in the above general theory and we also study these specific theories.

In section 3, the dual gravity theory is analyzed.
We study the Hawking-Page transitions 
and the thermodynamical instability of charged Kerr-AdS black
holes, and reveal the phase structures for these black holes. 
A schematic of the resulting phase diagram is shown in Fig.\ \ref{fig:CRBH-phase}
and we found a gap between the Hawking-Page line and the
instability line.
We compare these phase diagrams with those of the gauge theory,
and find remarkable agreement between them (Fig.\ \ref{fig:CR-Com}).
Furthermore, we calculate the ratio of the effective actions
between these two theories and show that the ratio takes a value close to 3/4
even at a low temperature.
This quantitative result shows that the deconfinement phase 
of free $\CN =4$ SYM with chemical potentials well describes
the dual black hole.
section 4 is devoted to discussion.

\section{Large-$N$ gauge theory}
In this section, we study the thermodynamics of $\CN =4$ SYM theory 
with the $U(N)$ gauge group on $S^3$.
First, we summarize the symmetry of this theory 
and the spectrum of its fields.
The $SU(4)$ and $SO(4)$ groups arise as the R-symmetry of the $\CN =4$
supersymmetry and the rotational symmetry of $S^3$, respectively.
The symmetry group has a $U(1)^5$ Cartan subgroup, and we can consider 
a grand canonical ensemble with five chemical potentials.

Then, we derive a partition function with chemical potentials.
We see that the partition function is reduced to a matrix model of the
Polyakov loop by summing over gauge invariant states or by integrating all
the massive modes.
The distribution of the eigenvalues of the matrix model exhibits 
a phase transition from the uniform phase to the nonuniform phase
at some critical temperature.
The low-temperature phase has thermodynamical quantities of order one,
while the high-temperature phase has those of order $N^2$.
The phase transition line is depicted in the phase diagram.

We will find that interesting phenomena occur in the case 
when more than four chemical potentials are turned on.
In this case, the maximal chemical potential $\m_{\text{max}}$ appears,
below which the entire body of the transition line is contained in the 
phase diagram.
We will also study the bounds of the chemical potentials above which
the unitarity of the theory breaks down.

We also consider the theory with only R-symmetry chemical potentials
or only $SO(4)$ chemical potentials as specific cases.
The R-symmetry case has already been studied in \cite{YY} 
and we obtain the same result here.

\subsection{Symmetry of $\CN=4$ super Yang-Mills theory and chemical potentials}\label{N=4review}
The AdS boundary of a charged Kerr-AdS black hole 
has $S^3$ topology in global coordinates.
Therefore, we need to study $\CN =4$ SYM on $S^3$. 
The action is given 
by\footnote{We take the normalization of the generator
  $T^a$ of the gauge group as $\text{tr} (T^aT^b) = \f{1}{2}\d^{ab}$.}
\begin{multline}
  S=-\int d^4x\sqrt{-g}\,\text{tr}\bigg[ \f{1}{2}(F_{\m\n})^2 +
  (D_\m\phi_m)^2 +l^{-2}\phi_m^2
    + i\bar\lambda^A\G^\m D_\m\lambda_A\\
-\f{g^2}{2}[\phi_m,\phi_n]^2  - g\bar\lambda^A
  \G^m [\phi_m,\lambda_A ]\bigg],
\label{eq:action}
\end{multline}
and the background metric is
\begin{equation}
 ds^2 = g_{\mu\nu}dx^\mu dx^\nu=-dt^2 + l^2d\Omega_3^2\ ,
\label{eq:BG}
\end{equation}
where $\m= 0,1,2,3,\ m=1,2,\dots,6,\ A=1,\dots,4$, 
$F_{\mu\nu}=\partial_\mu A_\nu-\partial_\nu A_\mu+ig[A_\mu,A_\nu]$,
$A_\mu$ is the gauge field of $U(N)$. 
$\phi_m$ is a scalar field and $\lambda_A$,
which is originally a gaugino in the $\bold{\Bar{16}}$ representation 
of ten-dimensional type IIB 
supergravity,\footnote{One can see from Table~\ref{tab:Spec} 
that the spinor $\lambda_A$  
satisfies the unitarity condition of the superconformal 
algebra~\cite{Mi,KMMR}
\begin{align}
2\{S,Q\}&=E-J_1-J_2-Q_1-Q_2-Q_3 \geq 0, \notag \\
2\{\Bar{S},\Bar{Q}\}&=E-J_1+J_2-Q_1-Q_2+Q_3\geq 0, \notag
\end{align}
only when we choose the $\Bar{\bold{16}}$ representation for $\lambda_A$.
This choice is appropriate:\, since the $\bold{16}$ representation does not
satisfy this condition.
}
is a four-dimensional spinor in the $(\bold{2}, \bold{\bar 4}) 
+ (\bold{\bar 2},\bold{4})$
representation under $SO(1,3)\times SU(4)$.
All fields are adjoint representation of $U(N)$. 
The gauge covariant derivative is defined by 
$D_\mu = \nabla_\mu + ig[A_\mu,\ \cdot\ ]$. 
$l$ is the radius of $S^3$ and we set $l=1$  for simplicity. 
The mass term of the scalar field $(\CR /6)\phi_m^2=l^{-2}\phi_m^2$ is
needed to make the theory conformal invariant, where $\CR$ is the Ricci scalar
in (\ref{eq:BG}).
This action has two types of global symmetry. 
One of them is $R_t\times SO(4)$, which arises from the symmetry of
the background spacetime~(\ref{eq:BG}), 
where $R_t$ represents the time translation invariance. 
The other one is $SO(6)\simeq SU(4)$, which originates from the R-symmetry 
of $\CN =4$ supersymmetry.

The conserved charges are associated with commutative (Cartan)
subgroups of global symmetry $R_t\times SO(4)\times SU(4)$. 
Due to the time translation symmetry $R_t$, the Hamiltonian $\hat{H}$ is 
conserved. 
The $SO(4)$ group contains a $U(1)^2$ Cartan subgroup and 
we denote the associated charges as $\hat{J}_1$ and $\hat{J}_2$.
These charges represent angular momenta on $S^3$.
The $SU(4)$ group also contains a $U(1)^3$ Cartan subgroup and 
we will denote the associated charges as $\hat{Q}_a\ (a=1,2,3)$. 
Therefore, we can consider a grand canonical ensemble with 
five chemical potentials in SYM at a finite temperature. 
The grand canonical partition function is given by
\begin{equation}
 Z(\beta)=\text{Tr}\,\left[e^{-\beta(\hat{H}- \sum_{a=1}^3\mu_a \hat{Q}_a
-\Omega_1\hat{J}_1-\Omega_2\hat{J}_2)}\right]
\label{eq:PF}
\end{equation}
where $\mu_a$, $\Omega_1$ and $\Omega_2$ are the chemical potentials 
conjugate to $\hat{Q}_a$,
$\hat{J}_1$ and $\hat{J}_2$, respectively. 
To calculate this partition function, we need
to know the eigenvalues of the conserved charges, 
$\hat{H}, \hat{Q}_a, \hat{J}_1$ and $\hat{J}_2$.

\subsection{Spectrum of conserved charges}\label{subsec:Spec}
First let us determine the R-charges of the fields using the method in
\cite{YY}.
The vector field is invariant under the $SO(6)$ group, thus has no R-charge.
When we write the six scalar fields as three complex fields
\begin{align}
  \Phi_1 \equiv \f{1}{\s2}(\phi_1 + i\phi_2),\quad 
  \Phi_2 \equiv \f{1}{\s2}(\phi_3 + i\phi_4),\quad
  \Phi_3 \equiv \f{1}{\s2}(\phi_5 + i\phi_6),
\end{align}
the generators $Q_i\ (i=1,2,3)$ of the Cartan subalgebra $U(1)^3$ of the
R-symmetry $SO(6)$ act on the complex vector 
\begin{align}
  \vec\Phi =(\Phi _{1},\Phi ^{\ast }_{1},\Phi _{2},\Phi ^{\ast }_{2},\Phi _{3},
  \Phi ^{\ast }_{3})^{T}
\end{align}
as rotations
\begin{align}
  Q^{\mathbf{6}}_{1}&=\text{diag}(1,-1,0,0,0,0)\ ,\notag \\ 
  Q^{\mathbf{6}}_{2}&=\text{diag}(0,0,1,-1,0,0)\ , \notag \\ 
  Q^{\mathbf{6}}_{3}&=\text{diag}(0,0,0,0,1,-1)\ . \label{RCscalar}
\end{align}

Four Weyl fermions $\lambda^A$ transform as fundamental representation 
$\mathbf{4}$ under $SU(4)_{R}$. 
We choose to represent the generators of the Cartan subalgebra, in the 
fundamental representation $\mathbf{4}$, following \cite{YY}, as
\begin{align}
  Q^{\mathbf {4}}_{1}&=\frac{1}{2}\text{diag}(1,1,-1,-1)\ , \notag \\
  Q^{\mathbf {4}}_{2}&=\frac{1}{2}\text{diag}(1,-1,1,-1)\ , \notag \\
  Q^{\mathbf {4}}_{3}&=\frac{1}{2}\text{diag}(1,-1,-1,1)\ . 
\end{align}
This choice is consistent with the assignment of the R-charges on the
scalar fields (\ref{RCscalar}), so that the antisymmetric representation 
$\mathbf{6}$ can be constructed from the tensor representation 
$\mathbf{4}\otimes \mathbf{4}$. 
Similarly, four conjugate Weyl fermions $\bar\lambda_{A}$ with the
representation $\bold{\bar 4}$ have the R-charges
\begin{align}
  Q^{\mathbf {\bar 4}}_{1}&=-\frac{1}{2}\text{diag}(1,1,-1,-1)\ , \notag \\
  Q^{\mathbf {\bar 4}}_{2}&=-\frac{1}{2}\text{diag}(1,-1,1,-1)\ , \notag \\
  Q^{\mathbf {\bar 4}}_{3}&=-\frac{1}{2}\text{diag}(1,-1,-1,1)\ . 
\end{align}

We now move on to the charges associated with the rotational group $SO(4)$.
We denote the generators of $SO(4)$ as $\boldsymbol{\hat J}_1$ and 
$\boldsymbol{\hat J}_2$ ($(\boldsymbol{\hat J}_1)_3 \equiv \hat J_1,
(\boldsymbol{\hat J}_2)_3\equiv \hat J_2$ ),
which satisfy the following commutation relation:
\begin{align}
  \{(\boldsymbol{\hat J}_1)_i, (\boldsymbol{\hat J}_1)_j\} &= i\ep_{ijk}(\boldsymbol{\hat J}_1)_k,\no
  \{(\boldsymbol{\hat J}_1)_i, (\boldsymbol{\hat J}_2)_j\} &= i\ep_{ijk}(\boldsymbol{\hat J}_2)_k, \notag \\
  \{(\boldsymbol{\hat J}_2)_i, (\boldsymbol{\hat J}_2)_j\} &= i\ep_{ijk}(\boldsymbol{\hat J}_1)_k. \qquad
  (i,j,k = 1,2,3) 
\end{align} 
$SO(4)$ can be represented
as two independent $SU(2)$ spins as
\begin{equation}
 \boldsymbol{\hat J}_1 = \boldsymbol{\hat j}_L + \boldsymbol{\hat j}_R,\qquad
 \boldsymbol{\hat J}_2 = \boldsymbol{\hat j}_L - \boldsymbol{\hat j}_R,
\label{eq:LK}
\end{equation}
where $\boldsymbol{\hat j}_L$ and $\boldsymbol{\hat j}_R$ represent the
generators of  two $SU(2)$ groups respectively, which satisfy
\begin{align}
  \{(\boldsymbol{\hat j}_L)_i, (\boldsymbol{\hat j}_L)_j\} &= i\ep_{ijk}(\boldsymbol{\hat j}_L)_k,\no
  \{(\boldsymbol{\hat j}_R)_i, (\boldsymbol{\hat j}_R)_j\} &= i\ep_{ijk}(\boldsymbol{\hat j}_R)_k,\notag \\
  \{(\boldsymbol{\hat j}_L)_i, (\boldsymbol{\hat j}_R)_j\} &= 0. 
\end{align}
All the fields on $S^3$ are characterized by the eigenvalues of 
their spins $(j_L, j_R)$ under the two $SU(2)$ groups.
The representations of the form
$(j,j\pm s)~(j=0,1/2,1,\dots)$ describe particles of spin $s$ \cite{HR}.
The Laplacian on $S^3$ and the Casimir operator
are related to each other. The relations 
for scalar fields $\phi$, spinor fields $\psi$ and 
divergenceless vector fields $A_i$ are given by
\begin{gather}
2(\, \boldsymbol{\hat j}_L^2 + \boldsymbol{\hat j}_R^2)\phi=-\nabla_{S^3}^2\phi\ ,\label{Casimir_sc}\\
2(\, \boldsymbol{\hat j}_L^2 + \boldsymbol{\hat j}_R^2)\psi=\left(-\nabla_{S^3}^2+\frac{\mathcal{R}}{8}\right)\psi\ ,\label{Casimir_sp}\\
2(\, \boldsymbol{\hat j}_L^2 + \boldsymbol{\hat j}_R^2) A_i=\left(-\nabla_{S^3}^2+\frac{\mathcal{R}}{3}\right) A_i\ ,\label{Casimir_v}
\end{gather}
where $\mathcal{R} = 6/l^2=6$ is the Ricci scalar of the three-sphere
and $\nabla_{S^3}^2$ is the Laplacian on $S^3$.
The operations of $\boldsymbol{\hat j}_L$ and
$\boldsymbol{\hat j}_R$ are defined by the Lie derivative along 
the $SU(2)$ generators. 
The proof of these relations is given in Appendix \ref{ap:SpLie}.

We now evaluate the spectrum of the conformally coupled
scalar with the representation $(j,j)$.
The equation of motion is
\begin{align}\label{CoSc}
\left[ \p_t^2 - \nabla_{S^3}^2 + \f{\mathcal{R}}{6}\right] \phi = 0.
\end{align}
From (\ref{Casimir_sc}) and (\ref{CoSc}), 
the energy spectrum for a scalar field is given by
\begin{equation}
\begin{split}
  E_s^2 &\equiv -\partial_t^2 = - \nabla_{S^3}^2 + 1\\
&=2(\, \boldsymbol{\hat j}_L^2 + \boldsymbol{\hat j}_R^2)+1=(2j+1)^2
,\qquad j=0,~\f12,~1,\dots .
\end{split}
\end{equation}
The degeneracy of the state with $(j,j)$ is $(2j+1)^2$.

Next we move on to the analogous calculation for Dirac fermions
represented as two Majorana fermions $(j,j+1/2) + (j+1/2,j)$.
The equation of motion for fermions is
\begin{equation}
 \left[ \p_t^2 - \Slash\nabla^2\right] \psi = 0,
\label{CoSp}
\end{equation}
where $\Slash\nabla\equiv \Gamma^i\nabla_i$ and $x^i$ are coordinates on $S^3$.
The spinor Laplacian is obtained from the square of the Dirac
 operator \cite{BD}
\begin{align}
  -\Slash\nabla^2 = -\nabla_{S^3}^2 + \f{\mathcal{R}}{4}.
\label{DtoL}
\end{align}
This is conformally covariant and we do not need the extra coupling to
the Ricci scalar in (\ref{CoSc}).
From (\ref{Casimir_sp}), (\ref{CoSp}) and (\ref{DtoL}), 
the energy of the fermion $(j,j+1/2)$ (or $(j+1/2,j)$) is
\begin{equation}
\begin{split}
  E_f^2 &= -\Slash\nabla^2
=2(\, \boldsymbol{\hat j}_L^2 + \boldsymbol{\hat
 j}_R^2)+\frac{3}{4}=\left(2j+\frac{3}{2}\right)^2\ ,
\end{split}
\end{equation}
and its degeneracy is $(2j+1)(2j+2)$.

Finally, we consider the divergenceless vector field
with the representation $(j,j+1) + (j+1,j)$. 
The equation of motion for the divergenceless vector is
\begin{equation}
 \left[ \p_t^2 - \nabla_{S^3}^2 + \f{\mathcal{R}}{3}\right] A_i = 0.
\label{CoV}
\end{equation} 
From (\ref{Casimir_v}) and (\ref{CoV}),
the energy spectrum for the vector field becomes
\begin{align}
  E_v^2 = - \nabla_{S^3}^2 + 2 = 2(\, \boldsymbol{\hat j}_L^2 +
 \boldsymbol{\hat j}_R^2) = (2j+2)^2\ ,
\end{align}
and the degeneracy is $(2j+1)(2j+3)$.
We summarize these spectra and the R-charges of free fields in
Table~\ref{tab:Spec}.
\begin{table}[htbp]
\hspace*{-0.4cm}
    \begin{tabular}{|c|c|c|c|c|}\hline
      Field &  $E$ & Degeneracy & Representation & $(Q_1,Q_2,Q_3)$\\
      \hline\hline
      scalar & $2j + 1$ & $(2j + 1)^2$ & $(j,j)$ & $(\pm 1,0,0)$, 
      $(0,\pm 1,0)$, $(0,0,\pm 1)$ \\ \hline
      \begin{minipage}{2cm}
        \centering
        fermion
        $(\bold{2}, \bold{\bar 4})$
      \end{minipage} & $2j + \f32$ & $(2j+1)(2j+2)$ & $ (j+\f12, j)$ & 
      \begin{minipage}{5cm}
\centering
        \vspace*{0.2cm}
      $(\f{1}{2},\f{1}{2},-\f{1}{2})$,\quad
      $(\f{1}{2},-\f{1}{2},\f{1}{2})$,\\
      $(-\f{1}{2},\f{1}{2},\f{1}{2})$,
      $(-\f{1}{2},-\f{1}{2},-\f{1}{2})$
      \vspace{0.25cm}
      \end{minipage}
      \\ \hline
      \begin{minipage}{2cm}
        \centering
        fermion
        $(\bold{\bar 2}, \bold{4})$
      \end{minipage} & $2j + \f32$ & $(2j+1)(2j+2)$ & $ (j, j+\f12)$ & 
      \begin{minipage}{5cm}
        \vspace*{0.2cm}
        \centering
      $(\f{1}{2},\f{1}{2},\f{1}{2})$,\quad  $(\f{1}{2},-\f{1}{2},-\f{1}{2})$,\\
      $(-\f{1}{2},\f{1}{2},-\f{1}{2})$,\ $(-\f{1}{2},-\f{1}{2},\f{1}{2})$
      \vspace{0.25cm}
      \end{minipage}
      \\ \hline
      vector & $2j +2$ & $(2j+1)(2j+3)$ & 
        $(j,j+1) + (j+1,j)$ & $(0,0,0)$ \\\hline 
    \end{tabular}
    \caption{Spectrum and R-charges of free fields. 
      The angular momentum $j$ has 
      half-integer values $0,~\f12,~1,\dots$.}
    \label{tab:Spec}
\end{table}

\subsection{Thermodynamics of $\mathcal{N}=4$ super Yang-Mills theory
  and phase transition} \label{YMTD}
Consider {\it free} Yang-Mills theory with {\it an arbitrary gauge group
  and matter} on {\it any compact space} with {\it any  chemical potential}
at a finite temperature.
In a compact space, all modes of the matter fields in the gauge theory are
massive and only the  zero modes of the temporal gauge field remain.
In this case, an exact expression for the partition function 
is given as follows: \cite{Su,AMMPV}
\begin{align}\label{parti3}
  Z(x) &= \text{Tr} \left[ e^{-\b (\hat H- \sum_i\tilde{\mu}_i\hat{N}_i)} \right],\\
  &= \int_G [dU]\exp\left\{ \sum_R \sum_{n=1}^\infty \f 1 n 
    [z_B^R(x^n,\tilde{\mu}_i) + (-)^{n+1}z_F^R(x^n,\tilde{\mu}_i)]\chi_R(U^n) \right\},\nonumber
\end{align}
where we denote the gauge group as $G$, its element as $U$,
the  character $\chi_R$ for the representation $R$.
The $\hat{N}_i$ are conserved charges and $\tilde{\mu}_i$ are chemical
potentials. 
We define single-particle partition functions of the boson and fermion 
for each representation $R$ as
\begin{align}\label{SPFs}
  z_B^R(x,\tilde{\mu}_i) \equiv \text{Tr}_{R}\ x^{\hat H-
    \sum_i\tilde{\mu}_i\hat{N}_i},\qquad
  z_F^R(x,\tilde{\mu}_i) \equiv
  \text{Tr}_{R}\ x^{\hat H- \sum_i\tilde{\mu}_i\hat{N}_i},
\end{align}
where $x=e^{-\b}$. 

We focus on $\mathcal{N}=4$ SYM on $S^1 \times S^{3}$ with
the gauge group $U(N)$. In this case, all matters are in the adjoint
representation of the gauge group and we only carry out summation over $R=adj$
in the partition function (\ref{parti3}). Then, the character
in~(\ref{SPFs}) becomes 
$\chi
_{adj}(U)=\chi_{fund}(U)\,\chi_{fund}(U^{\dagger})=\text{tr}(U)\,
\text{tr}(U^{\dagger })$.
The $U(N)$ group Haar measure $[dU]$ is given in Appendix~\ref{Ap:A}. 
The single-particle partition function (\ref{SPFs}) with chemical
potentials for scalar fields becomes the sum
of the partition function for each representation $(j,j)$
with energy $E=2j+1$ and R-charges $(Q_1,Q_2,Q_3)$.
This summation is carried out over $j$ and the $Q_a$ listed 
in Table~\ref{tab:Spec}. Its explicit form is evaluated as follows:
\begin{align}\label{PFSCR}
  &z_S(x, \Om_1, \Om_2, \mu_1, \mu_2, \mu_3) = \text{Tr}_{\text{scalar}}\left[ x^{
    \hat H - \sum_{a=1}^3\m_a\hat Q_a - \sum_{i=1}^2
      \Om_i\hat J_i } \right]\no
&= \sum_{\text{scalar}}x^{-\sum_{a=1}^3\m_a\hat Q_a }\sum_{j=0,1/2,\dots}^\infty
  \sum_{m_L=-j}^j\sum_{m_R=-j}^j x^{2j+1 - \Om_1(m_L+m_R) - \Om_2
    (m_L-m_R)}\no
  &= \f{x(1-x^2)(x^{\mu_1}+x^{-\mu_1} + x^{\mu_2}+x^{-\mu_2} +
    x^{\mu_3} + x^{-\mu_3})
  }{(1-x^{1+\Om_1})(1-x^{1+\Om_2})(1-x^{1-\Om_1})(1-x^{1-\Om_2})}.
\end{align}
Here we use the relations
$\hat J_1 = (\boldsymbol{\hat j}_L + \boldsymbol{\hat j}_R)_3$ and 
$\hat J_2 =  (\boldsymbol{\hat j}_L - \boldsymbol{\hat j}_R)_3$.
The Majorana fermion modes form the representation $(j,j+1/2) +
(j+1/2, j)$ with energy $E=2j+3/2$. The single-particle partition function
becomes
\begin{align}\label{PFFCR}
  &z_F(x, \Om_1, \Om_2, \mu_1, \mu_2, \mu_3) \no
  &=  \sum_{\text{chiral}}x^{-\sum_{a=1}^3\m_a\hat Q_a } 
  \sum_{j=0,1/2,\dots}^\infty
  \sum_{m_L=-j-1/2}^{j+1/2}\sum_{m_R=-j}^j x^{2j+\f32 -
    \Om_1(m_L+m_R) - \Om_2(m_L-m_R)} +\ \big[\text{anti-chiral} \big]\no
  &=  \f{ x^{\f32}(x^{\f{\Om_+}{2}} +
    x^{-\f{\Om_+}{2}} - x\,(x^{\f{\Om_-}{2}} +
    x^{-\f{\Om_-}{2}}))(x^\f{\m_1-\m_2-\m_3}{2} + x^\f{-\m_1 +\m_2
      - \m_3}{2} + 
  x^\f{-\m_1-\m_2+\m_3}{2} +x^{\f{\m_1+\m_2+\m_3}{2}})}
  {(1-x^{1+\Om_1})(1-x^{1+\Om_2})(1-x^{1-\Om_1})(1-x^{1-\Om_2})}\no
  &\qquad\qquad +\ \big[ (\m_1,\m_2,\m_3,\Om_2)\to - (\m_1,\m_2,\m_3,\Om_2)\big],
\end{align}
where we denote $\Om_+ \equiv \Om_1 + \Om_2$ and $\Om_- \equiv \Om_1 -
\Om_2$.
The vector modes form the representation $(j,j+1) +
(j+1, j)$ with energy $E=2j+2$ and no R-charge.
The single-particle partition function becomes
\begin{align}\label{PFVCR}
  &z_V(x, \Om_1, \Om_2, \mu_1, \mu_2, \mu_3) \no
  &= \sum_{j=0,1/2,\dots}^\infty
  \sum_{m_L=-j-1}^{j+1}\sum_{m_R=-j}^j x^{2j+2 - \Om_1(m_L+m_R) - \Om_2
    (m_L-m_R)} +(\Om_2\rightarrow -\Om_2)\no
  &= \f{x^2(1 + x^2  -x^{1+\Om_1} -x^{1-\Om_1}
    - x^{1+\Om_2} - x^{1-\Om_2} + x^{\Om_1 + \Om_2} + x^{-\Om_1-\Om_2})}
  {(1-x^{1+\Om_1})(1-x^{1+\Om_2})(1-x^{1-\Om_1})(1-x^{1-\Om_2})} + 
(\Om_2\rightarrow -\Om_2).
\end{align}
If we set $\Om_i$ or all the chemical potentials to zero, 
expressions (\ref{PFSCR}-\ref{PFVCR}) precisely
reduce to the single-particle partition functions given in \cite{YY}
and \cite{AMMPV}, respectively.
We obtain the partition function as a unitary 
matrix model:\footnote{In the following calculation, we will omit the arguments
$\Omega_1,\Omega_2,\mu_1,\mu_2$ and $\mu_3$ to simplify the expressions.}
\begin{align}\label{UMM}
  Z(x)=\int [dU] \exp \left( \sum_{m=1}^{\infty
    }\frac{1}{m}(z_{B}(x^{m})+(-1)^{m+1}z_{F}(x^{m}))\,tr(U^m)tr(U^{\dagger m
    })\right) ,
\end{align}
where $z_B(x)=z_S(x)+z_V(x)$. This expression is also derived by
a path integral in Appendix~\ref{Ap:path}.

The partition function (\ref{UMM}) can be expressed by the eigenvalues
$\{ e^{i\alpha _{i}}\} $ ($-\pi <\alpha _{i}<\pi ,\ i=1,\dots ,N$) 
of $U$ after rewriting the Haar measure given in (\ref{ap:measure2}).
The final expression becomes 
\begin{align}\label{PFf}
  Z(x)=\int \prod_{i=1}^N d\alpha _{i} \exp \left( -\sum_{i\neq j}V(\alpha _{i}-\alpha
    _{j})\right) ,
\end{align}
where
\begin{align}
  V(\theta )=\log 2+\sum_{n=1}^{\infty
  }\frac{1}{n}(1-z_{B}(x^{n})-(-1)^{n+1}z_{F}(x^{n}))\cos (n\theta ).
\end{align}
In the large-$N$ limit, the density of the eigenvalues becomes a continuous
function $\rho(\t)$. It must be nonnegative everywhere on 
$\{ -\pi < \t < \pi \}$ and can be normalized as 
$\int^\pi_{-\pi} d\t \>\! \rho(\t) =1$. 
The effective action $I_{\text{gauge}}\equiv -\ln Z$  of (\ref{PFf}) for $\rho(\t)$ becomes
\begin{align}\label{EffAct}
  I_{\text{gauge}}[\rho(\t)] = N^2\int d\t_1 \int d\t_2 ~\rho (\t_1) \rho
  (\t_2) V (\t_1 - \t_2) 
  = N^2 \sum_{n=1}^\infty \rho_n^2 V_n\ ,
\end{align}
where we define $\rho_n \equiv \int^\pi_{-\pi} d\t \>\! \rho(\t) \cos
(n\t)$ and 
\begin{align}
V_n \equiv \f{1}{\pi} \int^\pi_{-\pi} d\t \; V(\t) \cos (n\t)
= \f1n (1- z_B(x^n) - (-)^{n+1}z_F(x^n)).
\end{align}
This shows that the uniform eigenvalue distribution $\rho_n=0$ is an
absolute minimum when the inequality
\begin{align}
V_n>0
\quad \Leftrightarrow \quad  
z_B(x^n) + (-)^{n+1}z_F(x^n) < 1\qquad \text{for all}\;\; n
\end{align}
is satisfied.
Since the single-particle partition functions increase monotonically 
with $x$, and $x$ takes values $0 < x <1$, the condition of $n=1$ gives 
the lowest upper bound of $x$ above which the uniform distribution 
does not give a minimum of $I_{\text{gauge}}$.
Therefore, the critical temperature $T_H$, which separates the uniform phase
and the non-uniform phase, is determined by
\begin{align}\label{PTline}
  z_B(x_H) + z_F(x_H) = 1,
\end{align}
where $x_H\equiv e^{-1/T_H}$. 
That is to say, the sign of the coefficient $V_1$ determines
whether or not the density of the eigenvalues is uniform.

Near the above critical line, the coefficients $V_{n\geq 2}$ are
positive, whereas $V_1$ is negative.
Hence, the configuration that gives minimal $I_{\text{gauge}}$ is realized at 
$\rho_1=1/2$ and $\rho_{n\ge 2} = 0$, and thus  $I_{\text{gauge}}$ in (\ref{EffAct}) becomes
$\mathcal{O}(N^2)$. 
Below the critical line, on the other hand, the configuration 
$\rho_{n\ge 1}=0$ minimizes $I_{\text{gauge}}$. In this case $I_{\text{gauge}}$ becomes
$\mathcal{O}(1)$. 
Therefore, this phase transition is a confinement/deconfinement transition
of gauge theory: the phase of $T>T_H$ is the deconfinement
phase and that of $T<T_H$ is the confinement phase. 
We will solve this equation and reveal the phase structure in section \ref{PS}.

In \cite{Su,AMMPV}, the exact solution for $T>T_H$ is obtained in the
large-$N$ limit, while one can approximate this solution as follows if 
$z_n(x) \equiv z_B(x^n) + (-)^{n+1}z_F(x^n)$ 
decreases exponentially with $n$ for $n>1$:
\begin{gather}\label{ApDenFunc}
  \rho(\t) = 
  \begin{cases}
    \s{\sin^2\left(\f{\t_0}{2}\right) - \sin^2\left(\f\t
        2\right)} \cos\f{\t}{2} 
    \> \big/ \>
    \pi\sin^2\left(\f{\t_0}{2}\right)
    & \left(\left|\t\right|<\t_0\right)
    \\ 
    \qquad \qquad \qquad \qquad \;\;  0 
    & \left( \text{elsewhere} \right)
  \end{cases} \ \
  \\
  \sin^2\left(\f{\t_0}{2}\right) = 1- \s{1-\f{1}{z_1(x)}}.
\end{gather}
The factor $z_n$ does in fact decrease exponentially, 
and thus we can use (\ref{ApDenFunc}) as a good approximation.
Substituting (\ref{ApDenFunc}) into (\ref{EffAct}), we obtain the
effective action in a very simple form:
\begin{align}\label{actCFT}
  I_{\text{gauge}}= -N^2\left( \f{1}{2\sin^2 \left( \f{\t_0}{2} \right)} +
    \f{1}{2}\log \left(  \sin^2\left( \f{\t_0}{2} 
      \right) \right) -\f{1}{2} \right) .
\end{align}
For $T>T_H\ (z_1>1)$, this action is well-defined and exhibits a
first-order transition of $\mathcal{O} (N^2)$, 
while the action is zero for $T<T_H$ since all $\rho_n$ must be zero.
We can calculate this effective action once the
single-particle partition functions (\ref{SPFs}) are given.
It will be compared quantitatively with that of
dual gravity in section \ref{sc:ratio}.

\subsection{Phase structure}\label{PS}

\begin{figure}[htbp]
  \centering
  \subfigure[$\mu_1\equiv\mu, \mu_2=\mu_3=0$]
  {\includegraphics[width=5.2cm, clip]{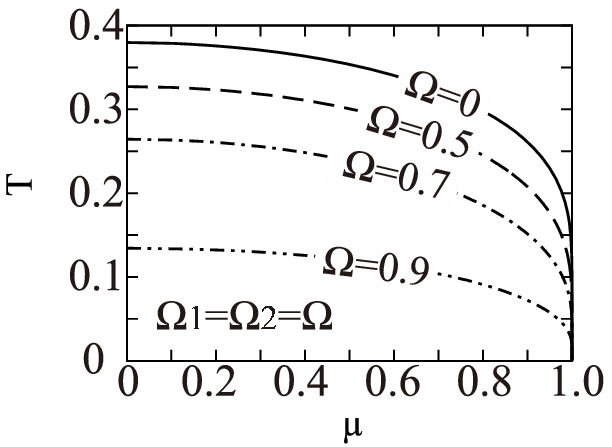}
    \label{YMCLP1}
  }
  \subfigure[$\mu_1=\mu_2\equiv\mu, \mu_3=0$]
  {\includegraphics[width=5.2cm, clip]{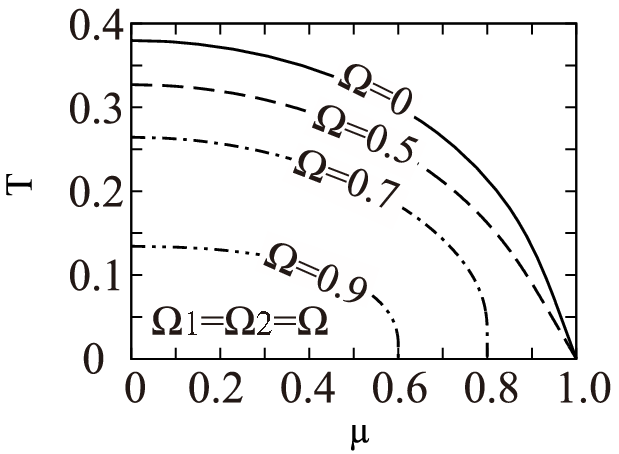} 
    \label{YMCLP2}
  }
  \subfigure[$\mu_1=\mu_2=\mu_3\equiv\mu$]
  {\includegraphics[width=5.2cm, clip]{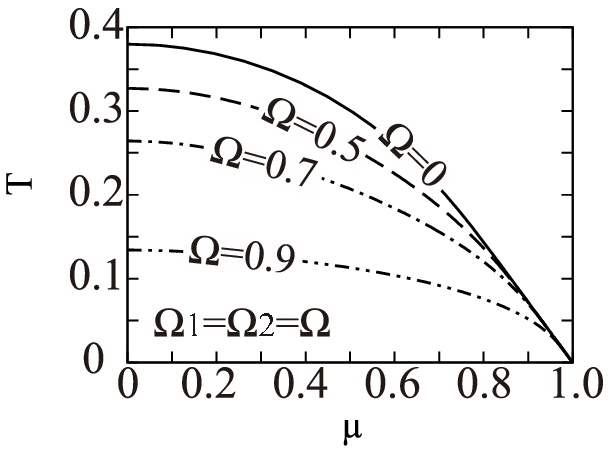}
    \label{YMCLP3}
  }
  \subfigure[$\m_a=0$]
  {\includegraphics[width=5.2cm, clip]{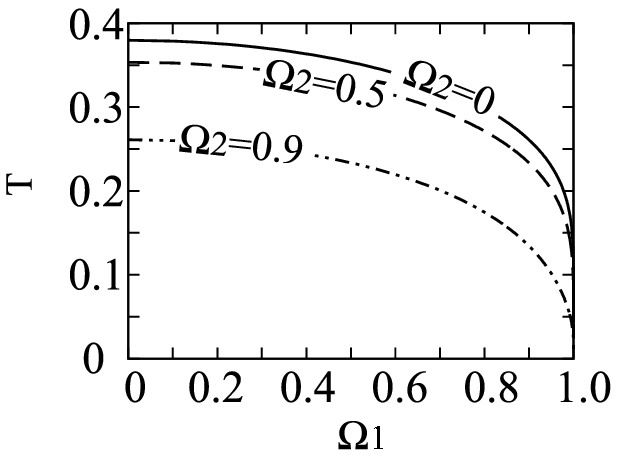}
    \label{fig:KN}
  }
  \subfigure[$\Om_i=0$]
  {\includegraphics[width=5.2cm, clip]{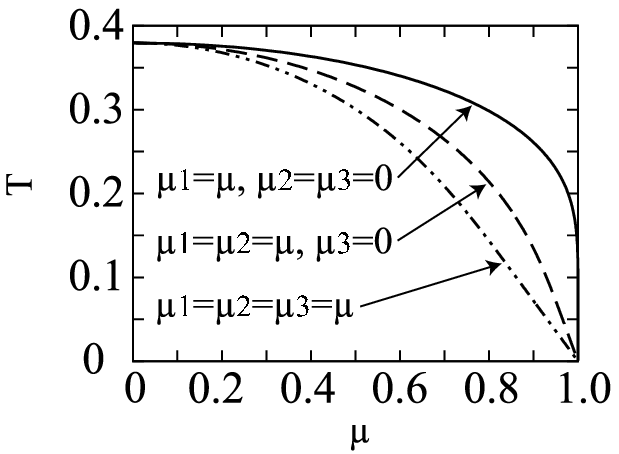}
    \label{fig:RN}
  } 
  \caption{
    Phase diagrams of $\CN =4$ large-$N$ SYM theory with 
    R-symmetry and $SO(4)$-symmetry chemical potentials.
    We plot the critical temperature $T_H$ for nonzero R-charge chemical 
    potentials $(\mu_1, \mu_2, \mu_3)$ in the cases of $\Omega_1=\Omega_2\equiv
    \Omega$ and $\Omega=0, 0.5, 0.7$ and $0.9$ in (a), (b) and (c).
    The confinement phase is below the line and 
    the deconfinement phase is above it.
    We set the R-charge chemical potentials as 
    (a) $(\mu, 0, 0)$,  
    (b) $(\mu, \mu, 0)$ and (c) $(\mu, \mu, \mu)$.  
    (d) shows the case that all $\mu_a$ are zero and two
    $\Om_i$ are unequal. In (e) all $\Om_i$ are zero and 
    only the R-charge chemical potential is turned on.
    (e) is equivalent to Fig.\ 2 of \cite{YY}.
  }
  \label{fig:CR}
\end{figure}

In the previous subsection, we derived Eq.~(\ref{PTline}) 
for the critical temperature of the phase transition. 
In this section, we solve this
equation numerically and depict the transition lines on 
the phase space. The phase space is the six-dimensional space of 
$(T,\Om_1,\Om_2,\mu_1, \mu_2, \mu_3)$ and we cannot cover the whole
phase space. We thus focus on several slices, which are 
$(\mu_1,\mu_2,\mu_3)=(\mu,0,0),(\mu,\mu,0),(\mu,\mu,\mu)$ 
and $\mu,\Omega_1,\Omega_2>0$. The confinement/deconfinement phase transition
lines of these slices are depicted in Fig.\ \ref{fig:CR}.
The confinement phase is below the line and
the deconfinement phase is above it.
Fig.\ \ref{fig:CR} shows that the critical line becomes lower as the $SO(4)$
chemical potential increases 
irrespective of the values of the R-symmetry chemical potentials.
Almost all lines converge to the point where $T=0$ and $\mu =1$ for $(\mu,0,0)$ 
and $(\mu, \mu, \mu)$ (Fig.\ \ref{YMCLP1} and \ref{YMCLP3}), 
while the lines for $(\mu, \mu, 0)$ with large $\Om$ 
end at some maximal chemical potential $\m_{\text{max}}(\Om)$ (Fig.\ \ref{YMCLP2}).
These behaviors can be understood 
by specifying where the partition function 
diverges at zero temperature.
In the limit of $x \rightarrow 0$,
the bosonic partition functions (\ref{PFSCR}) and (\ref{PFVCR}) diverge
only when one of the chemical potentials approaches one. 
On the other hand, 
the fermionic partition function (\ref{PFFCR}) 
can also diverge at some maximal chemical potential less than one
if there are more than four chemical potentials.
The general conditions for the convergence of a partition function for
$x\rightarrow 0$ can be written as
\begin{gather}
 |\Omega_1|,\;|\Omega_2|,\;|\mu_1|,\;|\mu_2|,\;|\mu_3| < 1\ , \notag \\
 3-|\mu_1+\mu_2|+\mu_3-|\Omega_1+\Omega_2|>0\ , \notag \\
 3-\mu_1-|\mu_2+\mu_3|-|\Omega_1-\Omega_2|>0\ ,
\label{conv_cond}
\end{gather}
where we have assumed $\mu_1\geq\mu_2\geq\mu_3$ without loss of
generality. 
In the cases of $(\mu_1,\mu_2,\mu_3)=(\mu,0,0),\,(\mu,\mu,0)$ and $(\mu,\mu,\mu)$,
there exist maximum values of $\mu$
above which the inequalities (\ref{conv_cond}) are not satisfied.
These maximum $\mu$, 
which we denote as $\m_{\text{max}}^{(1)},\m_{\text{max}}^{(2)}$ and $\m_{\text{max}}^{(3)}$,
respectively, are given by
\begin{equation}
\begin{split}
  &\m_{\text{max}}^{(1)}=1 \ ,\quad
  \m_{\text{max}}^{(2)} = \min\left(\f{3 - \Om_1 - \Om_2}{2},\ 1 \right)\ ,\quad
  \m_{\text{max}}^{(3)} = 
      1- \f{|\Om_1 - \Om_2|}{3}\ \ .
\end{split}
\label{mu_max}
\end{equation}
These maximum values coincide with the end points of the
transition lines at $T=0$ in Fig.\ \ref{fig:CR}.

\subsection{Unitarity line}\label{sbsec:Unit}
In this section, we determine the unitarity line 
where the phase diagram is bounded.
In the presence of the chemical potentials, the time derivative 
in the Lagrangian shifts as
\begin{align}
  \p_0 \to \p_0 -i\left( \sum_{a=1}^3\m_a Q_a + \sum_{i=1}^2 \Om_i J_i
    \right).
\label{shift}
\end{align}
Then, the Hamiltonian is shifted as 
$H\rightarrow H -\sum_{a=1}^3\m_a Q_a - \sum_{i=1}^2 \Om_i J_i$,
and the chemical potentials are introduced into the path integral 
as explained in Appendix~\ref{Ap:path}.
By the replacement of the time derivative in (\ref{shift}), 
the mass of the scalar with the representation
$(E_s=2j+1,m_L,m_R,1,0,0),~|m_L|\le j, |m_R|\le j$ shifts as
\begin{equation}
  m_{\text{scalar}}^2=E_s^2=(2j+1)^2 \quad\to \quad m_{\text{scalar}}^2=E_s^2 - \left(\m_1
  + (\Om_1+\Om_2)m_L + (\Om_1 - \Om_2)m_R \right)^2 .
\end{equation}
The $j=0$ mode first becomes tachyonic as the chemical
potentials increase, and this gives the bound $\m_1 = 1$ above which 
the theory breaks down.
The scalar modes with $(\m_1,\m_2,\m_3)=(0,1,0)$ and $(0,0,1)$ also give 
the bounds $\m_2=1$ and $\m_3=1$, respectively.
Similarly, the $j=\infty$ mode also requires the upper
bound $\Om_i=1$. The $j=\infty$ mode of the vector field also 
becomes tachyonic for $\Omega_i>1$, and thus it imposes the same upper bound on 
$\Om_i$.
Note that although the fermionic single-particle partition function
diverges for $\m > \m_{\text{max}}$, the theory does not break down owing to
Pauli's exclusion principle, while the tachyonic boson causes the
theory to breakdown above the unitarity line.\footnote{
We thank H. Kawai for providing us with this interpretation.}

It is noteworthy that the $j=\infty$ mode first becomes 
tachyonic as each $\Omega_i$ increases
for the following reason. 
In dual gravity theory, a similar phenomenon occurs:
the $j=\infty$ mode on the Kerr-AdS black hole background first 
becomes unstable as we increase $\Om_i$.
This instability is called a superradiant instability,
which is caused by wave amplification via a mechanism 
similar to the Penrose process and by wave reflection 
due to the potential barrier of the AdS spacetime \cite{KLR,Cardoso,Kodama}.
This similarity with the $j=\infty$ mode suggests that 
the bound $\Omega_i<1$ in gauge theory may correspond to the bound for the
superradiant instability of Kerr-AdS black holes in dual gravity
theory \cite{HR}.

This unitarity line meets to the transition line at $T=0$ for many cases
as shown in Fig.\ \ref{fig:CR}, while in general  
a gap appears between these two lines when there are more than four.
chemical potentials, as shown in (\ref{conv_cond}) and (\ref{mu_max}).\footnote{
Note that in Fig.\ \ref{YMCLP3}, we set $\Om_1=\Om_2$, thus there is no gap.}
In section \ref{sc:inst} 
we will provide a dual description of this unitarity line, which
we think is the line representing the black hole instability,
and we find remarkable agreement between their behaviors.

\section{Comparison with dual gravity}

In this section, we briefly review the properties of the five-dimensional 
asymptotically AdS black hole, which is dual to the gauge theory we
have considered in the previous section. 
We study the Hawking-Page transition 
and the thermodynamical instability of charged Kerr-AdS black
holes, and reveal phase structures for these black holes. 
We compare the phase structure of the charged Kerr-AdS black hole with
that of the dual gauge theory.
We also compute the ratio of the effective actions
between the gauge theory and its gravity dual,
and show that the ratio is  close to the universal value of $3/4$
over a wide range of temperatures.

\subsection{Dual gravity theory}
The most general dual gravity solution (black hole) that is
asymptotically AdS spacetime is
expected to be constructed within five-dimensional maximal $SO(6)$-gauged
$\mathcal{N}=8$ supergravity, because this theory
arises from the reduction of type IIB supergravity on $S^5$.
$SO(6)$ has three $U(1)$ Cartan subgroups, therefore the black
hole solution can have three independent charges. 
Hence, we may concentrate on the $U(1)^3$ 
parts of $SO(6)$ and consider $U(1)^3$-gauged $\mathcal{N}=2$ 
five-dimensional supergravity. These $U(1)^3$ charges correspond to
the R-charges in dual gauge theory.

The Lagrangian for the relevant bosonic sector of the maximal
gauged supergravity in five dimensions is given by
\begin{align}
\CL = \CR \ast \bold{1} - \f{1}{2} \sum_{i=1}^2\ast d\varphi_i\we d\varphi_i  -
  \f{1}{2}\sum_{a=1}^3 X_a^{-2} \ast F^a\we F^a  
  + 4 \sum_{a=1}^3 X_a^{-1}\ast \bold{1}  + 
  F^1\we  F^2\we A^3,
\label{Lag}
\end{align}
where
\begin{align}
X_1= e^{-\f{1}{\s6}\varphi_1 -\f{1}{\s2} \varphi_2},\qquad
X_2= e^{-\f{1}{\s6}\varphi_1 +\f{1}{\s2} \varphi_2},\qquad
X_3 = e^{\f{2}{\s6}\varphi_1}.
\end{align}
The charges are given by the Gaussian integrals
\begin{align}
  Q_a = \f{1}{16\pi G_5} \int_{S^3} 
\left(X_a^{-2}\ast F^a - \f{1}{2}\ep_{abc}A^b\we A^c\right),
\end{align}
and the angular momenta are calculated from the Komar integral
\begin{align}
  J=\f{1}{16\pi G_5}\int_{S^3}\ast dK,
\end{align}
where $K$ is the Killing vector, that generates the rotational symmetry
$U(1)$ of spacetime. 
$G_5$ is Newton's constant in five dimensions and it can be written as 
 $1/G_5 = \pi^3/G_{10} = 2N^2/\pi$ by setting the AdS space
 radius $l$ to one.
In the following subsections, we will sketch some black hole solutions 
within this theory~(\ref{Lag}).

\subsection{Five-dimensional charged Kerr-AdS black hole} 
The most general solution of (\ref{Lag}) can have two independent
rotations and three independent charges. These are five degrees of
freedom excluding the mass parameter. 
We denote these charges as 
$(J_1,J_2,Q_1,Q_2,Q_3)$, where $J_1$ and $J_2$ are angular momenta and
$Q_1,Q_2$ and $Q_3$ are $U(1)$ charges.  
Unfortunately such a general solution has not yet been
discovered. 
The currently known charged Kerr-AdS
black holes have three or four degrees of freedom. 
The solutions with three degree of freedom are 
$(J_1,J_2,Q_1,Q_1,Q_1)$,$^{40,}$\footnote{
The thermodynamics of the solution $(J_1,J_1,Q_1,Q_1,Q_1)$ constructed
in \cite{CvLuP} were studied and the field theory dual was discussed in
\cite{KuLu}.}
$(J_1,J_2,Q_1,0,0)$\cite{CCLP2} and  
$(J_1,J_2,Q_1,Q_1,Q_3(J_1,J_2,Q_1))$\cite{CCLP3}. 
The solutions with four degrees of freedom are  
$(J_1,J_2,Q_1,Q_1,Q_3)$\cite{MP},
$(J_1,J_1,Q_1,Q_2,$ $Q_3)$\cite{CvLP,CGLP}
and $(J_1,J_2,Q_1,Q_2,Q_3)$ with one constraint and supersymmetry \cite{KLR2}.

In this paper, we focus on the solution
$(J_1,J_1,Q_1,Q_2,Q_3)$\cite{CvLP,CGLP}. 
The metric is given by
\begin{align}\label{CLP}
  ds^2 &=  -\f{Y-f_3}{r^4 H^{2/3}}dt^2 + \f{r^4
    H^{1/3}}{Y}dr^2 + r^2H^{1/3}d\Om_3^2 + 
  \f{f_1- r^6 H}{r^4H^{2/3}}(\sin^2\t d\phi_1 + \cos^2\t d\phi_2)^2 \no
  & \qquad - \f{2f_2}{r^4H^{2/3}}dt(\sin^2\t d\phi_1 + \cos^2\t d\phi_2),\\
  A^a &= \f{2}{r^2H_a}\left\{ s_ac_adt + a(c_as_bs_c -
    s_ac_bc_c)(\sin^2\t d\phi_1 + \cos^2\t d\phi_2) \right\},\no
  X_a &= H_a^{-1}H^{1/3}, 
\nonumber
\end{align}
where the indices $a, b$ and $c$ run through $1,2,3$, where 
$a\neq b\neq c\neq a$, and
\begin{align}
  H &= H_1H_2H_3,\qquad H_a = 1 + \f{2ms_a^2}{r^2},\no
  d\Om_3^2 &= d\t^2 + \sin^2\t d\phi_1^2 + \cos^2\t d\phi_2^2,\no
  s_a &= \sinh \d_a,\qquad c_a = \cosh \d_a,
\end{align}
and the functions $f_1,f_2,f_3$ and $Y$ are given by
\begin{align}
  f_1 &= r^6H + 2m a^2 r^2 + 4 m^2a^2\left[2\Big(\prod_a c_a -
  \prod_as_a\Big)\prod_bs_b - \sum_{a<b}s_a^2s_b^2 \right],\no
  f_2 &= 2m a\Big(\prod_a c_a - \prod_as_a\Big)r^2 + 4m^2a\prod_as_a,\no
  f_3 &= 2ma^2(1+r^2) + 4m^2a^2\left[2\Big(\prod_a c_a -
  \prod_as_a\Big)\prod_bs_b - \sum_{a<b}s_a^2s_b^2\right],\no
  Y &= f_3 + r^6H + r^4 - 2mr^2.
\end{align}
The inverse temperature, entropy, angular velocity and electric
potentials are given as
\begin{align}
  \b &= \f{2\pi\s{f_1(r_+)}}{3r_+^4 + 2(1+ 2m\sum_a
    s_a^2)r_+^2 + 4m^2\sum_{a < b}s_a^2s_b^2- 2m(1-a^2)}, \\
  S &= N^2\pi \s{f_1(r_+)}, \label{S_CLP}\\
  \Om &= \f{f_2(r_+)}{f_1(r_+)},\\ 
  \m_a &= \f{2m}{r_+^2+2ms_a^2}\left(
    s_ac_a + a\f{f_2(r_+)}{f_1(r_+)}(c_as_bs_c - s_ac_bc_c) \right),
\end{align}
where the outer horizon $r_+$ is defined as the largest root of the
function $Y(r)$. The conserved charges are
\begin{align}
  M = N^2\f{m(3+a^2+2\sum_is_i^2)}{2},\qquad J = N^2 ma\big(
  \prod_ac_a - \prod_as_a \big),\qquad
  Q_a = N^2 ms_ac_a.
\end{align}
The effective action is given by
\begin{equation}
 I_\text{gravity}=(M-TS-2\Omega J-\mu_1 Q_1-\mu_2 Q_2-\mu_3 Q_3)/T\ .
\label{I_CLP}
\end{equation}

The value of the effective action and the free energy $F\equiv TI_{\text{gravity}}$
of the black hole
are measured relative to the thermal AdS space without a black hole.
Therefore, when the sign of the effective action or the free energy is negative
(or positive), the black hole phase is stable (or unstable) against 
the thermal AdS phase. This phase transition 
is well known to be the Hawking-Page transition,
and the transition line is characterized as $I_\text{gravity}=0$.

\subsection{Five-dimensional Kerr-AdS black hole} 
The charged Kerr-AdS black hole (\ref{CLP}) considered in the previous section 
contains the Kerr-AdS black hole with equal two rotations but 
does not contain the one with two independent rotations.
Hence, here we separately treat  the Kerr-AdS black hole with two
independent rotations.

The five-dimensional Kerr-AdS black hole is defined by the following
metric:\cite{HHT,GPP}
\begin{align}
  ds^2 &= - \frac{\Delta_r}{\rho^2} \left(dt - \frac{a_1
      \sin^2\theta}{\Xi_1}d\phi_1 -
  \frac{a_2 \cos^2\theta}{\Xi_2} d\phi_2\right)^2 +
  \frac{\Delta_{\theta}\sin^2\theta}{\rho^2} \left(a_1 dt -
  \frac{(r^2+a_1^2)}{\Xi_1} d\phi_1\right)^2 \no
  & + \frac{\Delta_{\theta}\cos^2\theta}{\rho^2} \left(a_2 dt -
  \frac{(r^2+a_2^2)}{\Xi_2} d\phi_2\right)^2 + \frac{\rho^2}{\Delta_r} dr^2 +
  \frac{\rho^2}{\Delta_{\theta}} d\theta^2 
\nonumber
\\
  & + \frac{(1+r^2)}{r^2 \rho^2}
  \left ( a_1a_2 dt - \frac{a_2 (r^2+a_1^2) \sin^2\theta}{\Xi_1}d\phi_1
    - \frac{a_1 (r^2 + a_2^2) \cos^2 \theta}{\Xi_2} d\phi_2 \right )^2,
\label{K-metric}
\end{align}
where 
\begin{align}
  \Delta_r &= \frac{1}{r^2} (r^2 + a_1^2) (r^2 + a_2^2) (1 + r^2) - 2m,
  \nonumber \\
  \Delta_{\theta} &=  1 - a_1^2 \cos^2\theta - a_2^2 
    \sin^2\theta, \nonumber  \\
  \rho^2 &=  r^2 + a_1^2 \cos^2\theta + a_2^2 \sin^2\theta ,
  \nonumber \\
  \Xi_i &= 1-a_i^2. 
\end{align}
The scalar and gauge fields are given by $X_a=1$ and $A^a=0$, respectively.
This metric is nonsingular outside the horizon at $r=r_+$ defined by
the larger root of the equation $\D_r (r_+) =0$ provided $a_i^2<1~(i=1,2)$. 
To obtain the appropriate conformal boundary, 
we use the following coordinates:
\begin{align}
 T &= t, \no
\Xi_1 y^2 \sin^2\Theta &= (r^2 + a_1^2) \sin^2\theta, \no
\Xi_2 y^2 \cos^2\Theta &= (r^2 + a_2^2) \cos^2\theta, \no
\Phi_i &= \phi_i + a_i t,
\end{align} 
which are nonrotating at infinity.
Using these coordinates, 
the angular velocities become
\begin{align}\label{Omega}
 \Omega_i = \frac{a_i (1+r_+^2)}{r_+^2+a_i^2},
\end{align}
and the conformal boundary becomes $R_t \times S^3$:
\begin{align}
 ds^2 = -dT^2 + d\Theta^2 + \sin^2 \Theta d\Phi_1^2 + \cos^2
 \Theta d\Phi_2^2,
\end{align}
as  expected.
The inverse Hawking temperature is determined
to avoid a conical singularity of the metric as 
\begin{align}\label{beta}
  \b = \f{2\pi r_+ (r_+^2+a_1^2)(r_+^2+a_2^2)}{2r_+^6 +
    (1+a_1^2+a_2^2) r_+^4 - a_1^2 a_2^2}.
\end{align}
The action relative to pure AdS space is 
\begin{align}\label{KerrAct}
  I_\text{gravity}  
  = -\frac{N^2\beta (r_+^2+a_1^2)(r_+^2+a_2^2)(r_+^2-1)} {4 r_+^2
    (1-a_1^2)(1-a_2^2)}.
\end{align}
This action is only negative  for $r_+>1$, and then the Hawking-Page transition
takes place at $r_+=1$. 
The entropy of this black hole is given by$^{52,}$\footnote{The black hole 
entropy given in \cite{HHT} is different from that in \cite{GPP} 
up to $\f{\pi}{2}$.}
\begin{align}\label{KBHEn}
  S = N^2\f{\pi (r_+^2 + a_1^2)(r_+^2 + a_2^2)}{r_+ (1-a_1^2)(1-a_2^2)},
\end{align}
and the mass and angular momenta are
\begin{align}\label{KBHMJ}
 M = N^2\frac{m(2\Xi_1 + 2\Xi_2 - \Xi_1\Xi_2)}{2\Xi_1^2\Xi_2^2}, \qquad 
 J_1 = N^2\frac{a_1 m}{\Xi_1^2\Xi_2},\qquad 
 J_2 = N^2\frac{a_2 m}{\Xi_1\Xi_2^2}. 
\end{align}

\subsection{Phase structure}\label{sc:GrPhase}
\begin{figure}[htbp]
  \centering
  \subfigure[$\mu_1\equiv\mu, \mu_2=\mu_3=0$]
  {\includegraphics[width=5.2cm, clip]{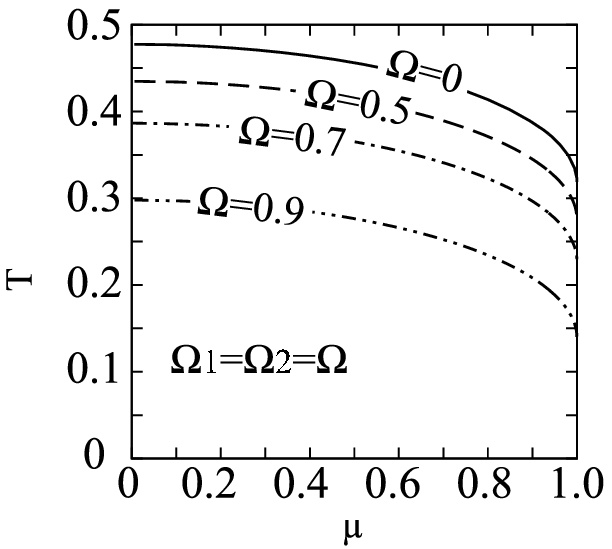}
    \label{CLP1}
  }
  \subfigure[$\mu_1=\mu_2\equiv\mu, \mu_3=0$]
  {\includegraphics[width=5.2cm, clip]{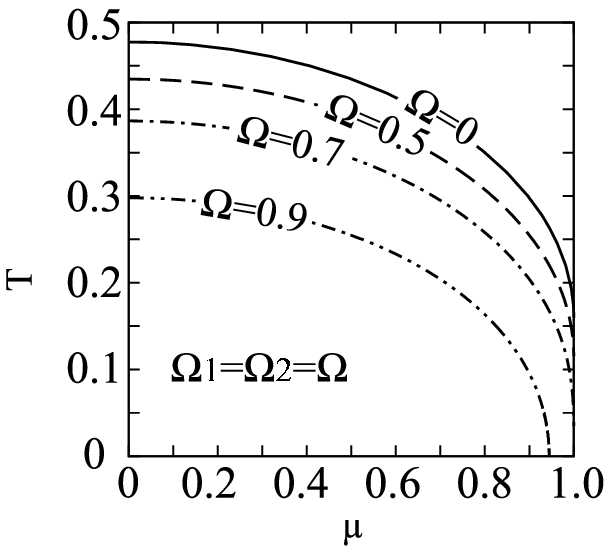} 
    \label{CLP2}
  }
  \subfigure[$\mu_1=\mu_2=\mu_3\equiv\mu$]
  {\includegraphics[width=5.2cm, clip]{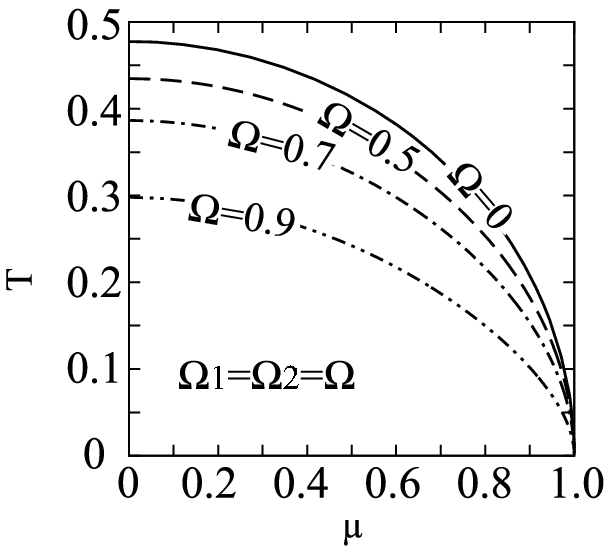}
    \label{CLP3}
  }
  \subfigure[Kerr-AdS BH]
  {\includegraphics[width=5.2cm, clip]{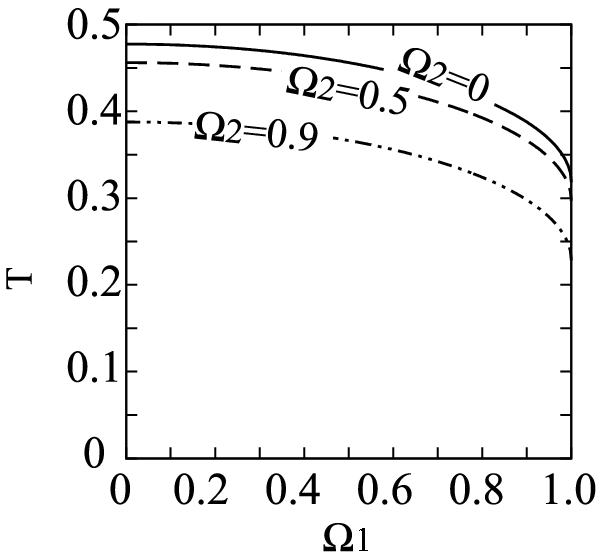}
    \label{fig:KerrBH}
  }
  \subfigure[R-charged BH]
  {\includegraphics[width=5.2cm, clip]{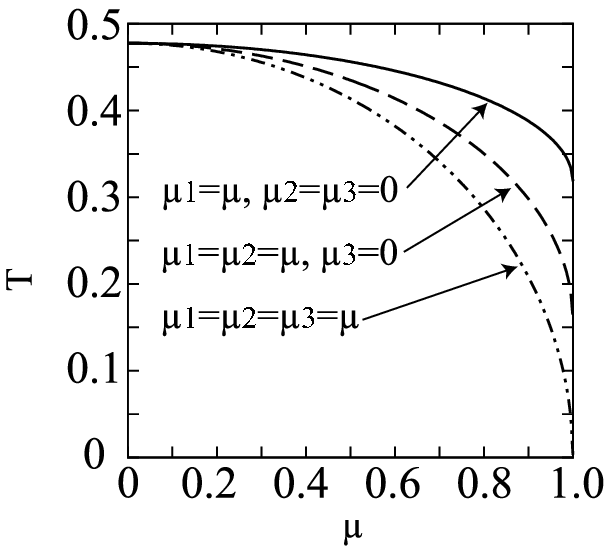}
    \label{fig:RNBH}
  }
  \caption{
    Phase diagrams for  charged Kerr-AdS black holes with two equal 
    rotations and three independent R-charges: $(J, J, Q_1,Q_2,Q_3)$. 
    We plot the transition lines for the R-symmetry chemical potentials
    (a) $(\m,0,0)$, (b) $(\m,\m,0)$ and (c) $(\m,\m,\m)$, varying the
    angular velocities $\Om=\Om_1=\Om_2$.
    We also depict the transition lines of Kerr-AdS and R-charged black holes.
    The lines represent the temperature of the Hawking-Page transition
    between the thermal AdS space and the Kerr-AdS black hole.
    The thermal AdS space is preferentially realized below the lines, 
    and the black hole is preferentially formed above the lines.
    These figures are drawn in the same scale as the phase diagrams for
    the gauge theory in Fig.\ \ref{fig:CR}.}
  \label{fig:CLP}
\end{figure}

Now let us consider the phase structure of the charged Kerr-AdS black hole.
The transition temperature is determined by the condition
$I_\text{gravity}(T,\m_i, \Omega_i)=0$ for the action (\ref{I_CLP}).
This equation is too complicated to obtain $T$ analytically in terms of $\mu_a$ and
$\Omega_i$, except for the limiting case of vanishing electric charges
 (Kerr-AdS black hole case) or vanishing rotations (R-charged
black hole case).
We therefore plot the diagrams numerically, which are shown in Fig.\ \ref{fig:CLP}.

First, we consider the phase diagrams of the charged Kerr-AdS black 
hole (Figs.~\ref{fig:CLP}(a)-\ref{fig:CLP}(c)). These are similar to the
phase diagrams for the dual gauge theory (Figs.~\ref{fig:CR}(a)-\ref{fig:CR}(c)).
In particular,
we can see the strong agreement between
 Fig.\ \ref{YMCLP2} and Fig.\ \ref{CLP2}. 
In this case, for $\Omega\gtrsim 0.9$, 
the transition line ends at $(\mu, T)=(\mu_\text{max}, 0)$, where $\mu_\text{max}<1$.
This appearance of $\mu_\text{max}$ also occurs in the gauge theory
for $\Omega> 0.5$ (see Fig.\ \ref{YMCLP2}).
This similarity may be evidence for the AdS/CFT correspondence.

The similarities of the phase diagrams for the gravity and the gauge theory 
can also be seen for Kerr-AdS black holes (Fig.\ \ref{fig:KerrBH}) 
and R-charged black holes (Fig.\ \ref{fig:RNBH}), which can be obtained as 
the nonrotating limit of the charged Kerr-AdS black holes.
For R-charged black holes, 
we can reproduce the phase diagram already obtained in \cite{CG,YY,Cha}.

These similarities show that 
a global phase structure such as a confinement/ deconfinement transition 
does not depend on the coupling constant if we regard the gravity theory 
to be the strongly coupled gauge theory via AdS/CFT. 
Instead of these marked similarities, there are some differences
between the phase diagrams for these two theories.
The transition temperatures for the gravity theory are higher than those for the
gauge theory in all cases.
Furthermore, the transition lines in the gravity theory can end at
$\mu=1, T>0$, but those in the gauge theory  always end at $T=0$.
This discrepancy may be due to the strong-coupling effect;
 the classical gravity theory is considered to be dual to the gauge theory 
in the strong-coupling regime, whereas we used the free gauge theory
to calculate the effective action and other quantities in section 2.

\subsection{Instability of charged Kerr-AdS black hole}\label{sc:inst}

In section\ref{sbsec:Unit}, we studied the unitarity line
for gauge theory. 
On the basis of the analysis of R-charged black holes,\cite{CG,YY,Yama} 
it has been suggested that this unitarity line in the gauge theory 
corresponds to the thermodynamical instability line on the
phase diagram in  dual gravity theory.
It will be interesting to study the thermodynamical stability line of charged
Kerr-AdS black holes and compare it with the unitarity line in the gauge
theory.\footnote{We are grateful to D. Yamada for giving us advice on
  the computation.}

The thermodynamical stability of a system can be analyzed as follows.
Suppose that we have a system in thermal equilibrium,
and we consider a small deviation from the equilibrium state.
The second law of thermodynamics is then written as
\begin{equation}
 \delta M -T\delta S- 2\Omega\delta J - \mu_a \delta Q_a 
= \delta M - x_i \delta X_i
\leq 0\ ,
\label{Second}
\end{equation}
where we define $x_i=(T,\,2\Omega,\,\mu_a)$ and $X_i=(S,\,J,\,Q_a)$. 
If there is a deviation that satisfies this second law~(\ref{Second}), 
it implies that the system is unstable thermodynamically.
Therefore, the stability condition is stated as
\begin{equation}
0 \leq \delta M - x_i \delta X_i
 =\left(\frac{\partial M(X)}{\partial X_i}-x_i\right)\delta X_i
 + \frac{1}{2}\frac{\partial^2 M(X)}{\partial X_i \partial X_j}\delta
 X_i \delta X_j\ ,
\label{StabCond}
\end{equation}
where we have neglected $\mathcal{O}\left((\delta X_i{})^3\right)$ terms.
The $\mathcal{O}(\delta X_i)$ terms in (\ref{StabCond}) vanish owing to
 Maxwell's relations if the system is in thermal equilibrium. Thus, if 
$M_{ij}\equiv\partial^2 M(X)/\partial X_i \partial X_j$ is positive definite, 
the thermal equilibrium system is stable. 
However, the explicit expression of $M_{ij}$ becomes complicated
due to the derivatives of $M$ with respect to $X_i$. 
To simplify the analysis of $M_{ij}$,
it is convenient to use parameters $(r_+,a,m,s_1,s_2,s_3)$ instead of $X_i$,
 because the derivatives of $M$ with respect to these parameters are much simpler.
Not all these parameters 
are independent of each other because of the equation $Y(r_+)=0$. 
Thus we may eliminate the rotation parameter $a$ using the equation $Y(r_+)=0$, 
and use the parameters $y_i\equiv(r_+,m,s_1,s_2,s_3)$. 
To use these parameters, it is convenient to define
\begin{equation}
 \tilde{F}(X,\tilde{x}) \equiv M(X) - \tilde{x}_i X_i\ ,
\end{equation}
 where $\tilde{x}_i$ are free parameters independent of $X_i$. 
This function $\tilde{F}$ is equal to 
the original Gibbs free energy for $\tilde{x}_i=x_i$,
and it satisfies
\begin{equation}
 \left.\frac{\partial \tilde{F}(X,\tilde{x})}{\partial
  X_i}\right|_{\tilde{x}=x}
=\left[\frac{\partial M(X)}{\partial
 X_i}-\tilde{x}_i\right]_{\tilde{x}=x}=0\ ,\qquad
 \left.\frac{\partial^2 \tilde{F}(X,\tilde{x})}
{\partial X_i \partial X_j}\right|_{\tilde{x}=x}
=\frac{\partial^2 M(X)}{\partial X_i \partial X_j}\ .
\label{tilFrel}
\end{equation}
Then the Hessian of $\tilde{F}$ becomes
\begin{equation}
\begin{split}
 H_{ij}&\equiv\left.\frac{\partial^2 \tilde{F}(X(y),\tilde{x})}
{\partial y_i \partial y_j}\right|_{\tilde{x}=x}
=\left.\frac{\partial X_l(y)}{\partial y_i} 
\frac{\partial^2 \tilde{F}(X,\tilde{x})}
{\partial X_l \partial X_k}\frac{\partial X_k(y)}{\partial y_j}
\right|_{\tilde{x}=x}
+\left.\frac{\partial^2 X_k(y)}{\partial y_i \partial y_j}
\frac{\partial \tilde{F}(X,\tilde{x})}{\partial
 X_k}\right|_{\tilde{x}=x}\\
&=\frac{\partial X_l(y)}{\partial y_i} 
\frac{\partial^2 M(X)}{\partial X_l \partial X_k}
\frac{\partial X_k(y)}{\partial y_j}\ .
\end{split}
\end{equation}
In the final equality we used (\ref{tilFrel}). 
Therefore, the positivity of $\det(M_{ij})$ is
equivalent to that of $\det\, (H_{ij})$ as long as $\p X_i(y)/\p y$
is nondegenerate, and thus the thermodynamical instability
occurs when $\det\, (H_{ij})=0$ and $\left|\p X_i(y)/\p y\right|\neq 0$.
This $H_{ij}$ is defined by the
derivatives with respect to the convenient parameters 
$y_i$ and it is easy to evaluate.

\begin{figure}[htbp]
  \centering
  \subfigure[$\mu_1\equiv\mu, \mu_2=\mu_3=0$]
  {\includegraphics[width=5cm,clip]{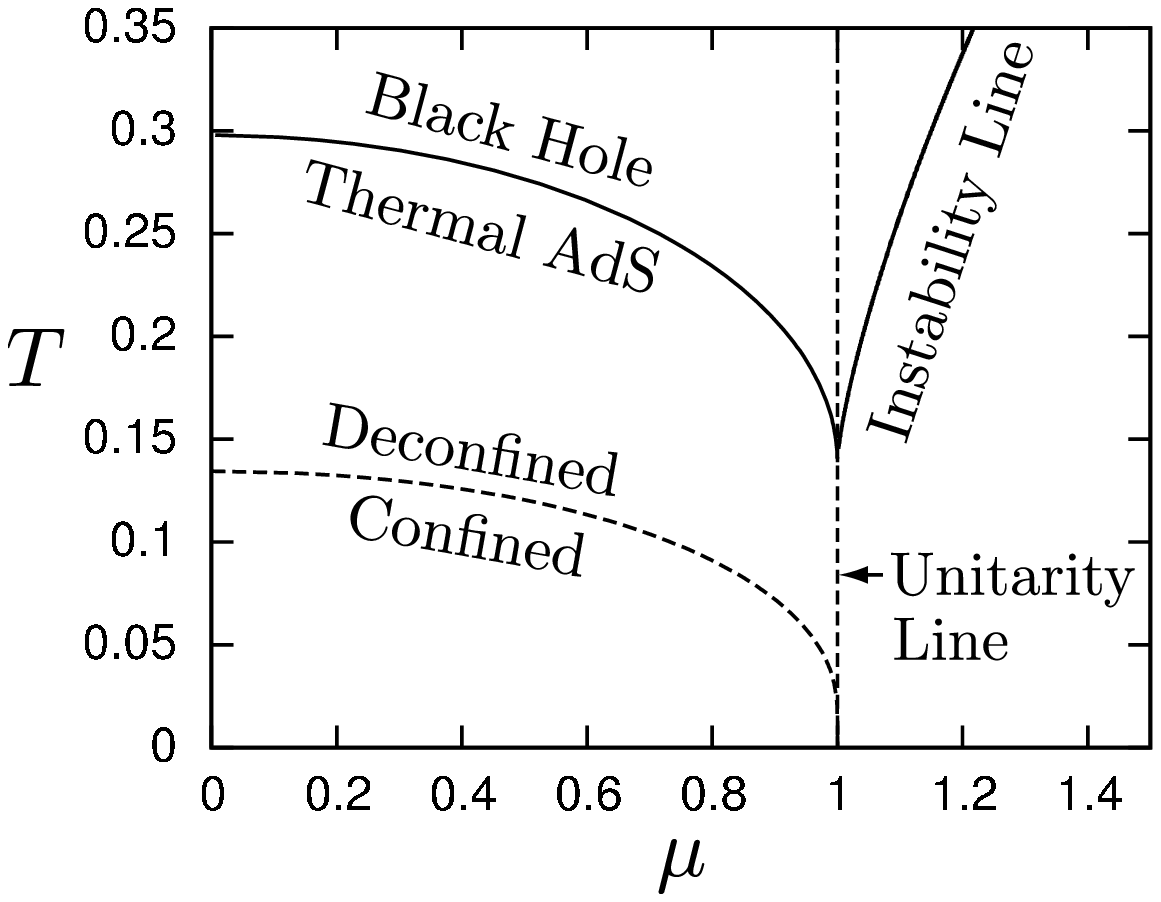}
    \label{uns1mu}
  }
  \subfigure[$\mu_1=\mu_2\equiv\mu, \mu_3=0$]
  {\includegraphics[width=5cm,clip]{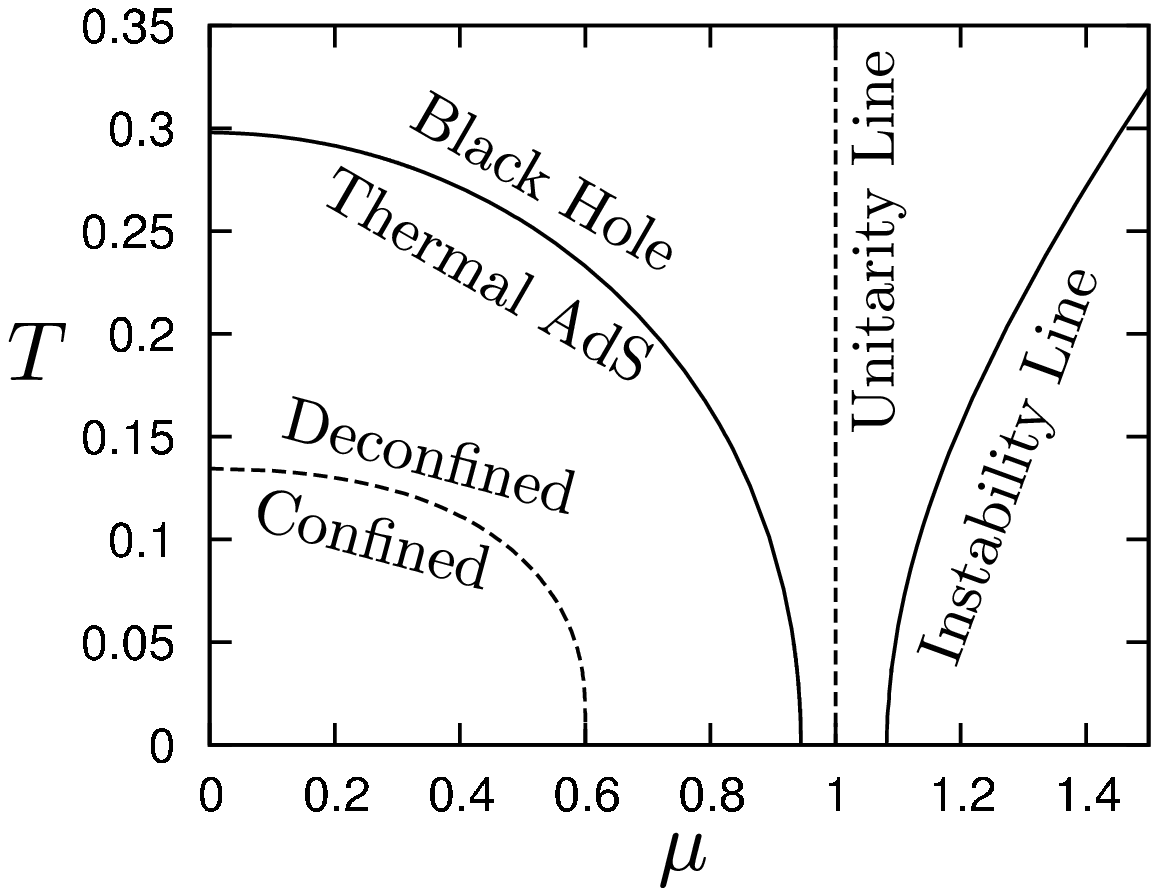} 
    \label{uns2mu}
  }
  \subfigure[$\mu_1=\mu_2=\mu_3\equiv\mu$]
  {\includegraphics[width=5cm,clip]{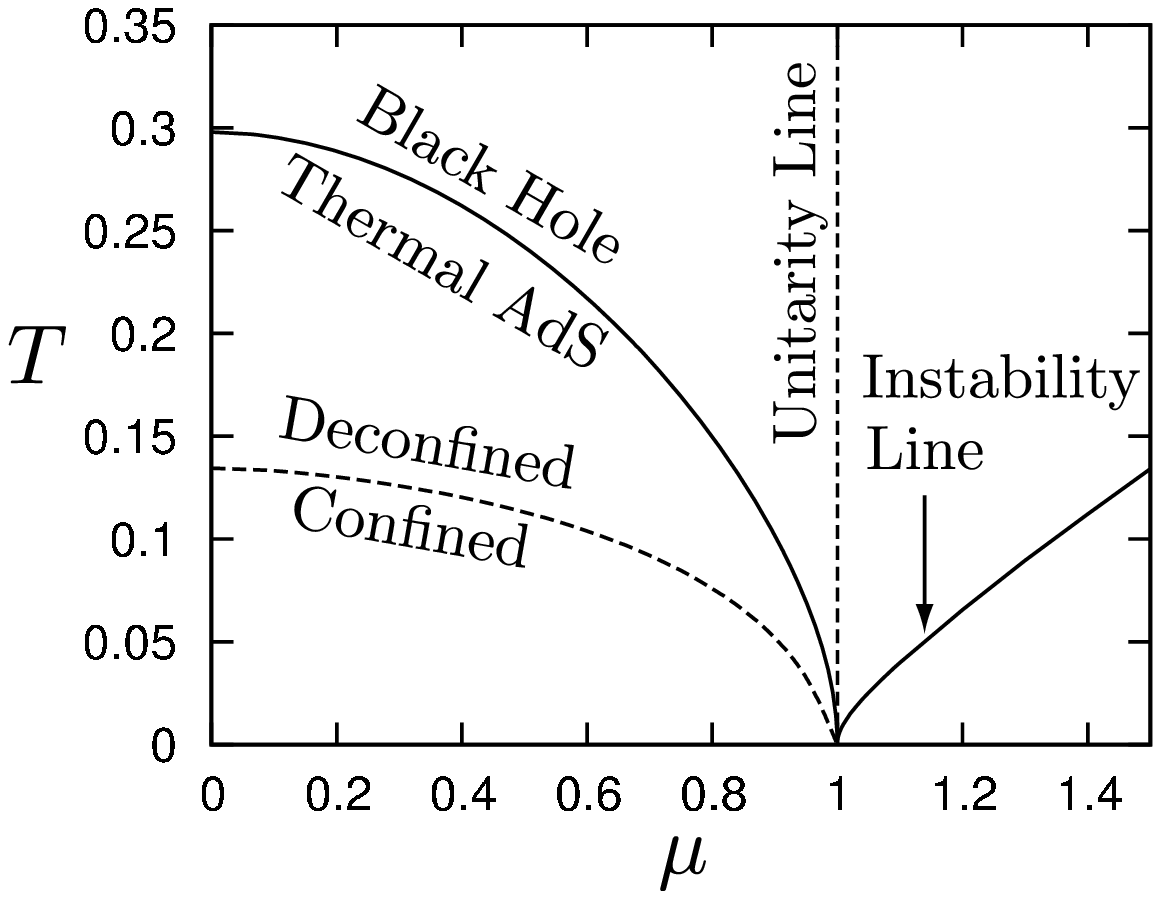}
    \label{uns3mu}
  }
  \caption{Phase diagrams for $\Omega=0.9$ including the
 unitarity lines. The solid lines are the Hawking-Page transition line and
 the instability line of a charged Kerr-AdS black hole.
 The dashed lines are the confinement/deconfinement transition line and
 the unitarity line of the dual gauge theory.}
  \label{fig:unstable}
\end{figure}

We evaluate $H_{ij}$ in the $(\mu,T)$ space
for the cases of 
$(\mu_1,\mu_2,\mu_3)=(\mu,0,0)$, $(\mu,\mu,0)$ and $(\mu,\mu,\mu)$,
fixing $\Omega=0.9$.\footnote{ 
We require a careful treatment for the evaluation of $H_{ij}$.
For example, in the case of $(\mu,\mu,\mu)$,
we have to evaluate $H_{ij}$ assuming that $s_1,s_2$ and $s_3$ are
independent of each other, and then
substitute $s_1=s_2=s_3$ to obtain the final result. 
Otherwise we cannot find the
instability in the cases $(\mu,\mu,0)$ and $(\mu,\mu,\mu)$ \cite{Yama}.}
Fig.\ \ref{fig:unstable} shows the resultant instability lines
on which $\det(H_{ij})$ becomes zero, along with the Hawking-Page
transition lines.
We also depict the confinement/deconfinement transition lines and the unitarity
lines of the dual gauge theory. 

In Fig.\ \ref{fig:unstable}, we can see qualitative similarities between
the instability line of the gravity theory and the unitarity line of
the dual gauge theory.
In particular, in  Fig.\ \ref{uns2mu} for the case 
$(\m_1, \m_2, \m_3)=(\mu,\mu,0)$, a gap appeared between the Hawking-Page line
and the instability line. 
We  found such a gap in the dual gauge theory
in section \ref{sbsec:Unit}.
At this point we can see a strong agreement between both theories.

It seems that the instability line corresponds to
the unitarity line in the gauge theory, whereas
there is a discrepancy between the gradients of the lines.
The instability line leans toward the large-$\mu$ region, while the unitarity
line is vertical at $\mu=1$.
This discrepancy can be resolved if we take a
quantum correction (nonzero gauge coupling) into account in the gauge theory. 
Actually, in the case of $\Om=0$, it has been shown in \cite{YY} that the
unitarity line is inclined and approaches the instability line of
the dual gravity as 't~Hooft coupling increases in a high-temperature regime.
The same behavior has also been found at a small finite temperature
in \cite{HKNW}.
Hence, we may expect that the unitarity line for $\Om\neq 0$ 
also begins to incline as
't~Hooft coupling increases and finally coincides with the
instability line in the strong-coupling limit.
Further investigations with nonzero gauge coupling are required
to confirm our prediction.

\subsection{Ratio of free energy at finite temperature} \label{sc:ratio}
To quantitatively study
the discrepancy between the free gauge theory and its dual gravity, we
evaluate the ratio of the effective actions as
\begin{align}
  f(T,\Omega_1,\Omega_2,\mu_1,\mu_2,\mu_3) \equiv
  \f{I_\text{gravity}}{I_\text{gauge}}\ ,
\label{ratio}
\end{align}
where the effective action of the free gauge theory is given 
by (\ref{actCFT}) and
$I_\text{gravity}$ is given by (\ref{I_CLP}) or (\ref{KerrAct}).
The ratio is plotted as a function of $T$ while fixing the chemical potentials. 

\begin{figure}[thb]
  \centering
  \subfigure[charged Kerr-AdS]{
    \begin{minipage}{6cm}
      \centering
        \includegraphics[height=4cm]{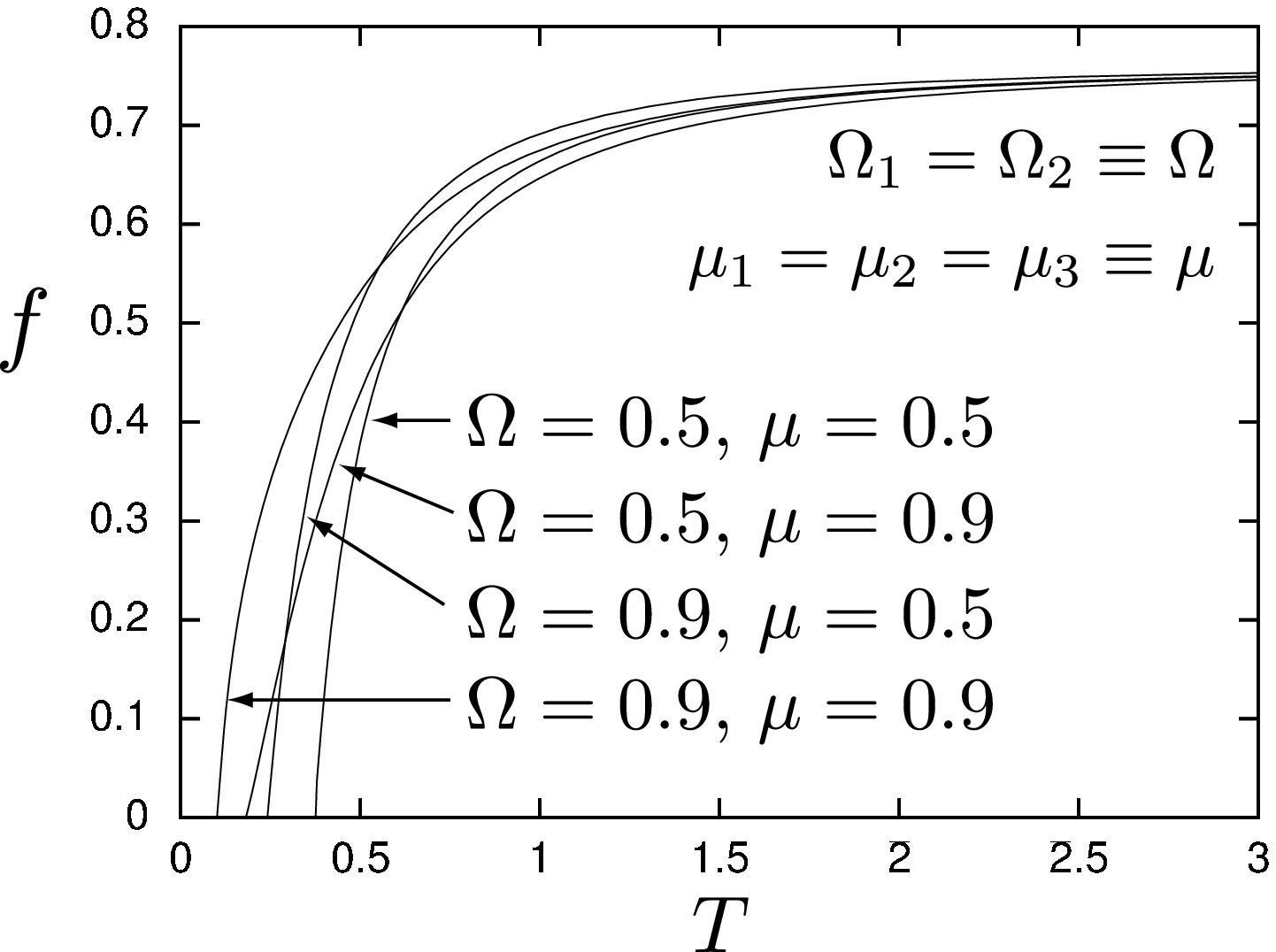}
        \label{CGLPRatio}
    \end{minipage}
  }
  \subfigure[Kerr-AdS]{
    \begin{minipage}{6cm}
      \centering
        \includegraphics[height=4cm]{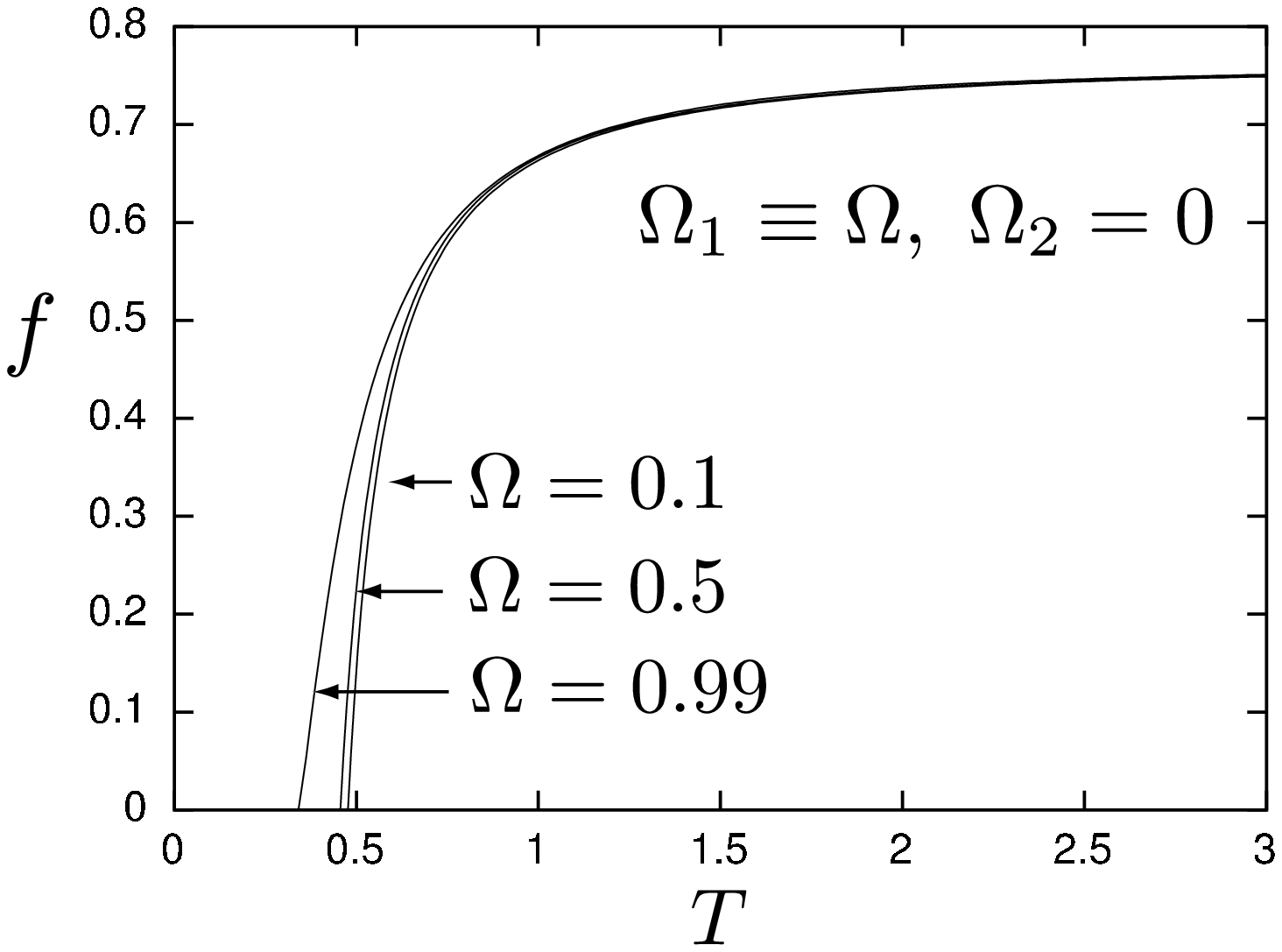}
        \label{KerrRatio}
    \end{minipage}
  }
  \caption{Ratio of the effective actions 
    for  charged Kerr-AdS black holes (a) and 
    Kerr-AdS black holes (b) to
    those for dual gauge theory as a function of temperature $T$. 
  }
\label{fig:ratio}
\end{figure}

We depict the ratios for the charged Kerr-AdS black holes for 
several values of $\Omega_1=\Omega_2\equiv \Omega$ and
$\m_1=\m_2=\m_3\equiv \mu$ 
in Fig.\ \ref{CGLPRatio}.
In Fig.\ \ref{KerrRatio}, we depict the ratios for the Kerr-AdS black holes
with unequal rotation: $\Omega_1 \neq 0$ and $ \Omega_2=0$.
We obtained the  ratios for other chemical potentials, such as a purely
R-charged case, which we do not show  here 
because their behaviors are similar to those of the cases above.
We find that the ratios  approach $3/4$ as the temperature increases
for any value of $\Omega_i$ or $\mu_a$.
Note that in the high-temperature limit, we cannot use the expression 
of the effective action for the gauge theory~(\ref{actCFT}), 
because (\ref{actCFT}) is valid only when $z_1>1$ and $z_{n\geq 2}<1$.
However, we can show analytically that the
ratio of the effective actions approaches 3/4 as $T\rightarrow \infty$ 
 using the Poincar\'{e} patch
in the limit where $S^3$ radius goes to infinity, as done in leading
order \cite{HR}. 
The subleading order was computed in \cite{LL}.

Surprisingly, the ratio remains approximately $3/4$ even at low temperatures.
This fact shows that the
gauge theory  corresponds fairly well to the dual charged Kerr-AdS black hole
even at low temperatures for any value of $\Omega_i$ and $\mu_a$.
However, the ratio becomes zero at some temperature in all cases.
This disagreement in the effective actions 
does not imply the breakdown of the duality;
it is merely due to the fact that the temperature of the Hawking-Page transition 
is always higher than that of the confinement/deconfinement
transition.

\section{Discussion}
We have studied the free $\CN =4$ SYM theory dual to the charged Kerr-AdS
black hole using the unitary matrix model.
We have seen that five chemical potentials can be introduced into the
thermodynamics of this theory; these chemical potentials 
are associated with the R-charges and the angular momenta.
We found the confinement/deconfinement transition and specified the
 unitarity bound for this theory. 
In the dual gravity theory, the Hawking-Page transition and the thermodynamical
instability of charged Kerr-AdS black holes have been investigated.
The resulting phase diagrams for gauge theory and charged Kerr-AdS black holes 
resemble each other, and, in particular, we have found that 
the confinement/deconfinement transition
line and the unitarity line of gauge theory correspond to 
the Hawking-Page transition line and the instability line in dual gravity
theory, respectively.
We have also found that  
the ratio of the effective actions of the two theories is always 3/4 at high
temperatures, and close to 3/4 even at low
temperatures around the Hawking-Page transition point,
for all values of the chemical potentials.
This result implies that the deconfinement phase 
of free $\CN =4$ SYM with chemical potentials describes 
the dual black hole well for all cases.

We have found interesting phenomena in  gauge theory and  dual gravity theory
when more than four chemical potentials are turned on.
In  gauge theory, 
the transition line touches the $T=0$ line at $\mu=\mu_{\text{max}}<1$
when the chemical potentials are set to
 $(\m_1,\m_2,\m_3, \Om_1, \Om_2)=(\mu,\mu,0,\Om,\Om)$ (Fig.\ \ref{fig:unstable}(b)).
In other words, a gap appears between the 
transition line and the unitarity line only for this case,
while the two lines touch at $T=0$ and $\mu=1$ 
for the other cases (Fig.\ \ref{fig:unstable}(a) and \ref{fig:unstable}(c)). 
The appearance of $\mu_{\text{max}}$ is caused by the divergence of the 
fermion partition function in the region $\mu>\mu_{\text{max}}$.
In dual gravity theory, on the other hand, 
a gap appears between the transition line and the instability line
only in the case  $(\m_1,\m_2,\m_3, \Om_1, \Om_2)=(\mu,\mu,0,\Om,\Om)$ 
(Fig.\ \ref{fig:unstable}(b)).
At this point the correspondence between these two theories is perfect.
However, the physical origin of this gap in the gravity theory is not
yet clear.
It will be interesting to investigate the origin and
the reason for the correspondence.

\begin{figure}[htbp]
  \centering
    \includegraphics[scale=0.5]{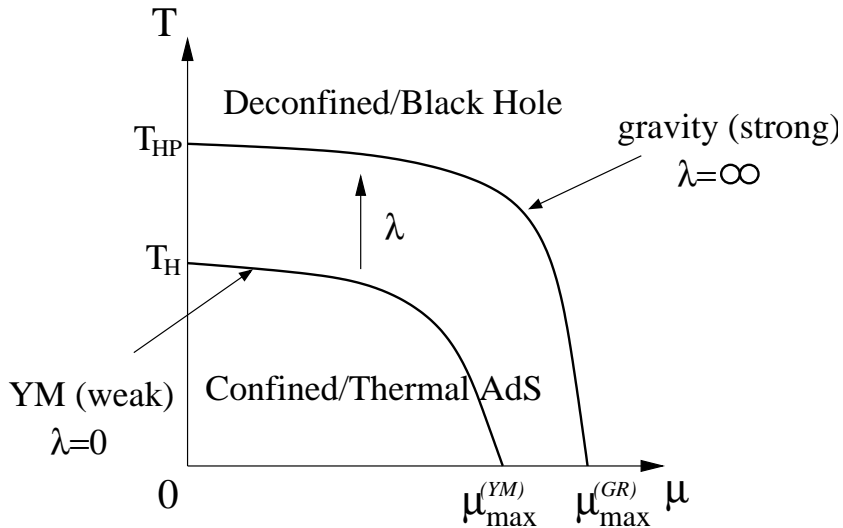}
  \caption{Plausible behavior of the phase transition line for 't~Hooft coupling.}
  \label{fig:Weak}
\end{figure}

Although the qualitative coincidence of 
gauge theory with chemical potentials and charged Kerr-AdS black holes
is good, we  have also found some discrepancies between these theories. 
First, the Hawking-Page transition temperature in the gravity theory
 is higher than the confinement/deconfinement transition temperature
 in dual gauge theory (Fig.\ \ref{fig:CR} and \ref{fig:CLP}).
Second, the instability lines incline toward the large-$\mu$ direction
for charged Kerr-AdS black holes, while the unitarity lines in dual gauge theory 
are vertical at $\mu=1$ (Fig.\ \ref{fig:unstable}). 
Finally, the ratio of effective actions is not one but almost $3/4$
(Fig.\ \ref{fig:ratio}). 
These discrepancies may be resolved if we consider  strong coupling
gauge theories, not the free theory we investigated in the literature. 
For the case of zero chemical potentials, there have been some works on 
finite gauge coupling effects.
It is known that the Hawking-Page transition temperature $T_{HP}$ decreases 
due to a string correction ($O(\al^3)$ correction) \cite{GKT,La}.
The $\al$ correction corresponds to the $1/\lambda$ correction in the 
strongly coupled gauge theory with gauge coupling $\lambda=\infty$;
thus, it is expected that the transition temperature monotonically
increases from $T_H$ at $\lambda=0$ to $T_{HP}$ at $\lambda=\infty$.
Weak coupling analysis of the transition temperature $T_H$ in
the gauge side has not yet been studied, but the coupling dependence
of $T_H$ in pure Yang-Mills theory was studied in \cite{AMMPR,AMR}, and it was 
shown that the temperature increases as the coupling becomes larger.
From these results we may expect that the phase transition line 
rises as the coupling $\lambda$ becomes large (see Fig.\ \ref{fig:Weak}).
In addition, in the case when the chemical potentials are zero,
the coupling constant dependence of the ratio of the effective actions 
is computed as \cite{GKT,KiRe,FoTa}
\begin{align}
  f(\lambda) = \f{I_{\text{gravity}}}{I_{\text{gauge}}(\lambda)} = \begin{cases}
    \f{3}{4} + \f{9}{8\pi^2}\lambda & \quad \lambda \sim 0, \\
    1 - \f{15}{8}\f{\zeta (3)}{(2\lambda)^{3/2}} & \quad \lambda \sim \infty,
    \end{cases}
\end{align}
where we used a stringy corrected gravity theory as a strongly coupled 
gauge theory.
This coupling dependence may not change even in the presence of
chemical potentials since the effect of chemical potentials is
negligible at high temperatures.
Recently, nonperturbative approaches to SYM theory have been developed
in \cite{AHNT,CaWi,ITT,IIST}, which are expected to clarify whether or not these 
discrepancies can be resolved beyond the zero coupling approximation.

It is interesting to compare our work with \cite{BLLM}, in which 
the gauge theory is analyzed by a hydrodynamic approach.
In this work the thermodynamical quantities of black holes 
were successfully reproduced from the gauge theory using a hydrodynamic approximation.
The results in \cite{BLLM} are in agreement with ours, at least in the case 
when this approximation is valid, \ie ~in the case that the temperature is 
sufficiently high.
Note that they treat the gauge theory in the strongly coupled
region as a perfect fluid, while we treat 
the gauge theory in the weakly coupled region as a free gas.
It is notable that in \cite{BLLM} the thermodynamical
quantities exhibit the same behavior as ours, although the regions of
 the gauge coupling studied are very different.

In this paper, we have used the black hole solutions with 
one degree of freedom in the angular momentum and three 
degrees of freedom in the R-charges. 
To check the correspondence for the most general case, we have
to use the solution with two independent rotations and
three independent charges, although such a solution has not been constructed yet.
We have found, however, that the parameter dependence of the black hole
solutions is regular and smooth. 
In addition, 
the behavior of the gauge theory is reasonably smooth for the 
chemical potential values even in the most general case.
Thus we expect that the properties of the most general 
black hole are similar to those we used in the literature.
It is of course desirable to check this directly using the most general exact 
solution, which is expected to be found in the future.
One prediction for this most general black hole solution is that a gap 
will appear between the transition line and the instability line
for the case $(\m, \m, \m, \Om_1, \Om_2)$, where $\Om_1\neq \Om_2$, for which
a gap appears in the dual gauge theory (see Eq.~(\ref{mu_max})).
It will be another nontrivial test of the correspondence to verify that
such a gap really appears in this case.

Another further investigation in the weak coupling analysis of field theory 
is to search for the phase transition dual to
the Gregory-Laflamme instability \cite{GrLa} in the gravity theory (see
Ref.~68 for a comprehensive review).
It has been shown that the black hole/black string transition 
corresponds to the phase transition in $1+1$ dimensional SYM on a
circle in \cite{AMMW}, and its generalization to a higher-dimensional
case was carried out in \cite{HO-GL,HN}.
Such a phase transition of $\CN =4$ SYM on $S^3$  was studied
in \cite{HKN} by computing an effective action at a finite temperature
and weak 't~Hooft coupling.
It was shown that the effective potential has a new saddle point
that preserves only an $SO(5)$ subgroup of the $SO(6)$ R-symmetry
above some critical temperature, which was identified as the
Gregory-Laflamme instability of the small AdS black hole
predicted in \cite{HuRa}.
It would be interesting to search for such a phase transition in
the presence of chemical potentials.

Our results suggest that the dynamical instability of AdS black holes can be
understood in terms of the unitarity violation in the gauge theory.
It is known that AdS black holes become unstable
due to the superradiant instability when their rotation is too fast.
Although the stability analysis of rotating black holes is difficult because of
the difficulty in separating the variables, there are some works on
this subject \cite{KLR,Cardoso,Kodama,MuSo2,MuSo}.
In these works, it was found that a Kerr-AdS black hole with
$\Omega_i>1$ suffers from the superradiant instability, and the modes with
a higher wave number first becomes unstable as $\Omega_i$ increases. 
In section \ref{sbsec:Unit}, we found that this behavior also appears 
in dual gauge theory, that is, higher modes of scalar and vector fields become
tachyonic when $\Omega_i>1$. Because of these tachyonic fields, 
the path integral diverges and the thermodynamical quantities cannot be defined.
However, if we take into account the
nonzero gauge coupling, the path integral will converge owing to the $\phi^4$ 
term and $A_\mu^4$ terms, and the thermodynamics will become well-defined. 
In this case the vector field $A_\m$ will acquire a nonzero vacuum expectation value
and will break the $SO(4)$ symmetry, which is the rotational symmetry 
of $S^3$. 
In gravity theory, on the other hand, the $U(1)^2$ rotational symmetry of spacetime 
is broken by the superradiant instability.
Thus, there is a possibility that the symmetry breaking vacuum in dual
gauge theory, which emerges due to the nonzero gauge coupling effect,
is the AdS/CFT counterpart of the final state of the black hole 
spacetime after the superradiant instability occurs.
We do not know much about such a
spacetime, thus it is interesting that we may be able to shed new light on 
this issue by analysis of  dual gauge theory.
This needs further investigation with nonzero gauge coupling and also
some analysis in  gravity theory to confirm our expectation.

The generalization to lower supersymmetry is an ambitious issue.
The $\CN =2$ case has already been done in \cite{LaOb} by considering the 
supersymmetric orbifold gauge theory dual to $AdS_5 \times S^5/\BZ_M$.
If we change $\CN =4$ SYM to $\CN =1$ SCFTs, the five-dimensional 
Newton constant will be modified in dual gravity reduced to five dimensions. 
It will be interesting to study the phase structure of $\CN =1$ SCFTs 
at zero coupling and verify the ratio of the thermodynamical quantities.
At high temperatures these ratios are known to always be approximately $3/4$,
\cite{NiTa} which gives quantitative evidence for
the AdS/CFT correspondence for $\CN =1$ SUSY.
Investigation in this direction would give further evidence 
for the correspondence at low temperatures.

\vspace{1cm}
\centerline{\bf Acknowledgements}

We are grateful to M. Fujita, M. Hanada, H. Hata, T. Hirata, H. Irie, 
T. Kikuchi, K. Matsumoto, S. Minakami, S. Minwalla, A. Miwa, Y. Sumitomo,
T. Tanaka, A. Tsuji and especially to
H. Kawai, T. Takayanagi and D. Yamada for fruitful discussions and 
important comments on the draft paper.

The work of KM and TN is supported by JSPS Grant-in-Aid for Scientific Research
No. 19$\cdot$3715 and No. 19$\cdot$3589, respectively.

\appendix

\section{Laplacian and Casimir operator on $S^3$}
\label{ap:SpLie}
In this appendix, we will prove of the relations
(\ref{Casimir_sc}), (\ref{Casimir_sp}) and (\ref{Casimir_v}).
Since these relations can be proved in similar ways, 
we will only show the relation for fermions (\ref{Casimir_sp}).
The left hand side of (\ref{Casimir_sp}) is interpreted as the spinor Lie 
derivative acting on fermions.
The spinor Lie derivative is defined as
\begin{equation}\label{SpLie}
  \CL_X \psi = X^i\nabla_i\psi - \f14 \nabla_i X_j \G^{ij}\psi.
\end{equation}
We have to take the Casimir operators $\boldsymbol{\hat j}_L$ and 
$\boldsymbol{\hat j}_R$ of a three-sphere for $X^i$ in this expression,
so let us start with the construction of their explicit forms.
The three-sphere is parameterized as\footnote{We use the
  parameterization in \cite{MuSo}.}
\begin{equation}\label{SU2metric}
  ds^2 = g_{ij}dx^i dx^j = \f14 ((\sigma^1)^2 + (\sigma^2)^2 + (\sigma^3)^2),
\end{equation}
where we have used the invariant forms $\sigma^a\ (a=1,2,3)$ of $SU(2)$
satisfying the relation $d\sigma^a = \f12
\ep^{abc}\sigma^b\we\sigma^c$ with 
\begin{equation}
  \sigma^1 = -\sin\chi d\t + \cos\chi\sin\t d\phi\ ,\ 
  \sigma^2 = \cos\chi d\t + \sin\chi\sin\t d\phi\ ,\ 
  \sigma^3 = d\chi + \cos\t d\phi\ .
\end{equation}
The coordinate ranges are $0\le \t < \pi,\ 0\le \phi < 2\pi$ and $0\le \chi
< 4\pi$.
We define the dual vectors $e_a$ of $\sigma^a$ by $\sigma^a_i e_b^i = \d_b^a$.
Using this relation and (\ref{SU2metric}), we obtain 
the following identity for the dual vectors:
\begin{align}\label{dualvec}
  \sum_{a=1}^3 e_a^i e_a^j = \f{g^{ij}}{4}.
\end{align}
The explicit forms of the dual vectors are 
\begin{gather}\label{Kilone}
  e_1 = -\sin\chi \p_\t + \f{\cos\chi}{\sin\t}\p_\phi - \cot\t\cos\chi \p_\chi\ ,\nonumber\\
  e_2 = \cos\chi \p_\t + \f{\sin\chi}{\sin\t}\p_\phi - \cot\t\sin\chi \p_\chi\ ,\quad
  e_3 = \p_\chi,
\end{gather}
which are in fact Killing vectors.
There is another set of Killing vectors $\xi_a\ (a=1,2,3)$,
\begin{gather}\label{Kiltwo}
  \xi_1 = \cos\phi \p_\t + \f{\sin\phi}{\sin\t}\p_\chi - \cot\t\sin\phi \p_\phi\ ,\nonumber\\
  \xi_2 = -\sin\phi \p_\t + \f{\cos\phi}{\sin\t}\p_\chi - \cot\t\cos\phi \p_\phi\ ,\quad
  \xi_3 = \p_\phi,
\end{gather}
which satisfy a similar identity to (\ref{dualvec}):
\begin{align}\label{Killvec}
  \sum_{a=1}^3 \xi_a^i \xi_a^j = \f{g^{ij}}{4}.
\end{align}
By an explicit calculation, we obtain the useful relation
\begin{equation}
 \sum_{a=1,2,3}\left(e^i_a \nabla_j e^k_a + \xi^i_a \nabla_j
		\xi^k_a\right)=0\ .
\label{US}
\end{equation}
We now introduce the following notation:
\begin{equation}
 (E^i_p)\equiv(e^i_1,e^i_2,e^i_3,\xi^i_1,\xi^i_2,\xi^i_3)\ .
\end{equation}
In this notation, we can express 
(\ref{dualvec}), (\ref{Killvec}) and (\ref{US}) as
\begin{align}
 E_p^i E_p^j = \f{g^{ij}}{2}\ ,\quad
 E^i_p \nabla_j E^k_p=0\ ,\label{US2}
\end{align}
where $\sum_{p=1}^6$ is omitted. 

The Casimir operators are related to the two sets of Killing vectors as
\begin{align}
  (\hat j_L)_a = i\xi_a,\qquad (\hat j_R)_a = -ie_a,
\end{align}
and the left hand side of (\ref{Casimir_sp}) becomes
\begin{align}
  2(\ \boldsymbol{\hat j}_L^2 + \boldsymbol{\hat j}_R^2) &=
 -2\mathcal{L}_{E_p}\mathcal{L}_{E_p}.
\label{Cas_Lie}
\end{align}
Using the definition of the Lie derivative for spinor (\ref{SpLie}), 
we can calculate the right hand side of (\ref{Cas_Lie}) as
\begin{align}
 \mathcal{L}_{E_p}\mathcal{L}_{E_p}\psi 
=&\,E^i_p E^k_p \nabla_i \nabla_k \psi
+E^i_p (\nabla_i E^k_p) \nabla_k \psi
-\frac{1}{4}E_p^i (\nabla_i\nabla_k E_l^p) \Gamma^{kl}\psi \notag \\
&\qquad -\frac{1}{2}E_p^i (\nabla_k E_l^p) \Gamma^{kl}\nabla_i\psi
+\frac{1}{16}(\nabla_i E_j^p)(\nabla_k E_l^p)\Gamma^{ij}\Gamma^{kl}\psi \notag\\
=&\, 
 \frac{1}{2} \nabla^2 \psi 
-\frac{1}{4}E^i_p\left(\nabla_i\nabla_kE^p_l\right)\Gamma^{kl}\psi
+\frac{1}{16}(\nabla_iE^p_j)(\nabla_kE^p_l)\Gamma^{ij}\Gamma^{kl}\psi\ .
\label{LLpsi}
\end{align}
In the last equality, (\ref{US2}) has been used. 
Here we need the following formulae to proceed further:
\begin{gather}
\nabla_i\nabla_kE_l^p=-\mathcal{R}_{kli}{}^jE_j^p\ ,\label{KilLem}\\
\Gamma^{ij}\Gamma^{kl}=\Gamma^{ijkl}
-2(g^{k[i}\Gamma^{j]l}-g^{l[i}\Gamma^{j]k})
-(g^{k[i}g^{j]l}-g^{l[i}g^{j]k})\ ,\label{GG}
\end{gather}
where $\mathcal{R}_{ijkl}$ is the Riemann tensor of $S^3$. 
For three-dimensional space, the $\Gamma^{ijkl}$ must vanish in (\ref{GG}). 
From (\ref{KilLem}), we obtain
\begin{equation}
 E^i_p\nabla_i\nabla_kE^p_l=-\mathcal{R}_{klij}E_p^iE_p^j=0.
\end{equation}
Thus, the second term in (\ref{LLpsi}) vanishes. 
From (\ref{GG}) and (\ref{US2}),
\begin{align}
(\nabla_iE^p_j)(\nabla_kE^p_l)\Gamma^{ij}\Gamma^{kl}
&=-4(\nabla_iE_j^p)(\nabla^i E_l^p)\Gamma^{jl} 
-2 (\nabla_i E_j^p)(\nabla^i E^j_p) \notag \\
&=-2 \nabla_i( E_j^p\nabla^i E^j_p)+2E_j^p \nabla^2 E^j_p \notag \\
&=-2\mathcal{R}_{jk}E^j_p E^k_p =-\mathcal{R}\ .
\end{align}
Therefore, the Casimir operator can be written as
\begin{equation}
 2(\ \boldsymbol{\hat j}_L^2 + \boldsymbol{\hat j}_R^2)\psi
=\left(-\nabla_{S^3}^2+\frac{\mathcal{R}}{8}\right)\psi\ .
\end{equation}

Similarly, we can obtain the relation of the Casimir operator and the Laplacian for
scalar and vector fields as
\begin{equation}
2(\ \boldsymbol{\hat j}_L^2 + \boldsymbol{\hat j}_R^2) \phi=-\nabla_{S^3}^2\, \phi\ ,\quad
2(\ \boldsymbol{\hat j}_L^2 + \boldsymbol{\hat
 j}_R^2)A_i=\left(-\nabla_{S^3}^2+\frac{\mathcal{R}}{3}\right) A_i\ .
\end{equation}

\section{Haar measure of $U(N)$}\label{Ap:A}

We define the metric of a unitary matrix as 
\begin{align}
  ||dU||^2 = \text{tr}(dUdU^\dagger),
\end{align}
where $U$ is an element of the unitary group $U(N)$ and we suppose that
$U$ depends on one parameter $t$.
It is clear that this metric is invariant under constant unitary
matrix rotations $U(t) \to VU(t)V^\dagger$.
The unitary matrix $U(t)$ can be diagonalized by some unitary matrix
$\Om(t)$, where
\begin{align}
  U(t) = \Om(t) M(t) \Om(t)^\dagger, \quad M = \text{diag}(e^{i\a_1},\dots,e^{i\a_N}),\quad
  \Om(t) \in U(N).
\end{align}
We express the unitary matrix $\Om(t)$ as $\Om(t) = e^{iT(t)}$
using the Hermite matrix $T(t)$.
Using the above decomposition, the metric becomes a separated form
\begin{align}\label{ap:measure}
  ||dU||^2 = \text{tr}(|dM|^2 + |[M, \Om^\dagger d\Om]|^2)
  = \sum_{i=1}^N (d\a_i)^2 + \sum_{i,j = 1}^N
  |e^{i\a_1}-e^{i\a_j}|^2|dT_{ij}|^2 ,
\end{align}
where we use $\text{tr}(dM^\dagger[M, \Om^\dagger d\Om])=
\text{tr}([dM^\dagger,M]\Om^\dagger d\Om) =0$.
The independent variables are $\a_i$, Re$T_{ij}$ and Im$T_{ij}$ for
$i<j$. Taking a basis of $(\a_i, \text{Re}T_{ij}, \text{Im}T_{ij})
~(i<j)$ and denoting $\lambda_{ij}=|e^{i\a_i} - e^{i\a_j}|$, the metric becomes
\begin{align}
  G = \left(
  \begin{array}{@{\,}ccc|ccc|ccc@{\,}}
    1 &  &  &  &  &  &  &  &  \\
     & \ddots  &  &  &   &  &  &  & \\
     &   & 1 &  &  &  &  &  & \\ \hline
      &   &  & \lambda_{12} &  &  &  &  & \\ 
      &   &  &  & \ddots &  &  &  & \\   
      &   &  &  &  & \lambda_{N-1,N} &  &  & \\ \hline
      &  &  &  &   &  & \lambda_{12} &  &   \\ 
      &  &  &  &  &  &  & \ddots &   \\   
      &  &  &  &  &  &  &  & \lambda_{N-1,N}    
  \end{array}
  \right).
\label{ap:measure2}
\end{align}
The Haar measure can be read from (\ref{ap:measure}):
\begin{align}
  [dU] &=\prod_{i=1}^N [d\a_i]\prod_{j<k}
  d\text{Re}T_{jk} d\text{Im}T_{jk} \cdot\s{\det G} \no
  &= \prod_{i=1}^N [d\a_i]\prod_{j<k}|e^{i\a_j} - e^{i\a_k}|^2
  d\text{Re}T_{jk} d\text{Im}T_{jk}\no
  &=  \prod_{i=1}^N [d\a_i]\prod_{j<k}4\sin^2\left(\f{\a_j - \a_k}{2}
  \right) [d\Om].
\end{align}
The term $[d\Om]$ is the gauge volume and should be divided when we
consider a gauge invariant action.

\section{Derivation of partition function with chemical potentials}\label{Ap:path}
In this appendix, we calculate the partition function~(\ref{eq:PF}) and
derive (\ref{UMM}). This partition function has been derived by a group
theoretical method in \cite{Su,AMMPV}. However, the path integral
derivation is also important because it will be a first step to take
into account the gauge coupling \cite{YY}.

Using (\ref{eq:LK}), we rewrite the partition function~(\ref{eq:PF}) as
\begin{equation}
 Z(\beta)=\text{Tr}\, \left[ e^{-\beta(\hat{H}-\mu_a \hat{Q}_a
-\Omega_+\hat{m}_L-\Omega_-\hat{m}_R)}\right]\ ,
\label{eq:PF3}
\end{equation}
where we define $(\, \boldsymbol{\hat j}_L)_3=\hat{m}_L$, 
$(\, \boldsymbol{\hat j}_R)_3\equiv\hat{m}_R$, $\Omega_+\equiv\Omega_1+\Omega_2$ and
$\Omega_-\equiv\Omega_1-\Omega_2$. 
It is well known that the partition function without chemical
potentials is given by the Euclidean path integral. 
To introduce nonzero chemical potentials, 
we need some tricks\cite{YY}: We replace $D_0$ by 
$D_0-i\mu_a \hat{Q}_a - i\Omega_+\hat{m}_L - i\Omega_-\hat{m}_R$
in the Lorentzian action~(\ref{eq:Lfree}).
Then the Hamiltonian is replaced by
$\hat{H}-\mu_a\hat{Q}_a-\Omega_+\hat{m}_L - \Omega_-\hat{m}_R$.
For the Euclidean signature case, we thus obtain
\begin{equation}
  D_0\rightarrow D_0-\mu_a\hat{Q}_a-\Omega_+\hat{m}_L -
   \Omega_-\hat{m}_R
\equiv D'_0\ .
\label{eq:D0'}
\end{equation} 

It is difficult to evaluate the partition function~(\ref{eq:PF3}) 
for finite gauge coupling; thus, we consider 
the free theory taking the limit of
$g\rightarrow 0$. We should take this limit carefully. 
Since we are considering field theory on a compact space,
the Gauss' law constraint becomes important. That is, the total $U(N)$
charge on the compact space should be neutral.
To take this into account, we decompose the gauge
field as
\begin{equation}
 A_0(x^\mu) = \tilde{A}_0(x^\mu) + \frac{1}{g}a(t)\ ,\quad
\frac{1}{g}a(t)\equiv \frac{1}{\omega_3}\int_{S^3}A_0(x^\mu)\ ,
\label{eq:A0dec}
\end{equation}
where $a(t)$ is the zero mode of $A_0$ and $\omega_3$ is the area of
the unit $S^3$. Then, the zero mode of $A_0$
becomes $\mathcal{O}(g^{-1})$ and $a(t)$ couples with other fields 
even in the limit of $g\rightarrow 0$. 
Using this device, in the limit of $g\rightarrow 0$, the 
SYM action~(\ref{eq:action}) of the Euclidean signature 
becomes
\begin{equation}
S= \int d^4x\sqrt{g}\ \text{tr}\left[\frac{1}{2}
F_{\mu\nu}F^{\mu\nu}+(D_\mu' \phi_m)^2 + l^{-2}(\phi_m)^2
+i\bar{\lambda}_A\Dslash'\lambda_A
\right]\ ,
\label{eq:Lfree}
\end{equation}
where the background metric is
\begin{equation}
  ds^2 = g_{\mu\nu}dx^\mu dx^\nu=d\tau^2 + l^2d\Omega_3^2
\end{equation}
and the differential operators and field strength are defined by
\begin{align}
&D_\mu'=(D_0',D_i)\ ,\quad
D_0=\partial_0 + i[a,\ \cdot\ ]\ ,\quad D_i=\nabla_i\ , \notag\\
&F_{0i}=D_0' A_i - \partial_i \tilde{A}_0\ ,\quad 
F_{ij}=\partial_i A_j - \partial_j A_i\ ,
\end{align}
where $\nabla_i$ is the covariant derivative on $S^3$ and $D_0'$ is defined in
(\ref{eq:D0'}). 
The finite temperature partition function~(\ref{eq:PF3}) can be written
as the Euclidean path integral,
\begin{equation}
 Z(\beta)=\int \mathcal{D}a\mathcal{D}\tilde{A}_0 \mathcal{D}A_i\mathcal{D}\phi \mathcal{D}\lambda
\,\exp(-S[\tilde{A}_0,A_i,\phi,\lambda,a])\ .
\end{equation}
The Euclidean time $\tau$ is periodic under $\tau\sim \tau + \beta$. 
The boson and fermion fields in the Euclidean action 
are periodic and antiperiodic under $\tau\sim \tau + \beta$,
respectively.

Now we take the Coulomb gauge for gauge fixing:
\begin{align}\label{Coulomb}
  \nabla_i A^i = 0.
\end{align}
This leaves one remaining degree of gauge freedom $a(\tau) \to a(\tau) + D_0'u_0(\tau)$,
where $u_0(\tau)$ is an arbitrary function consistent with the periodicity. 
We fix this as%
\footnote{We cannot take the gauge condition $a(\tau)=0$, because
it does not make $u_0(\tau)$ periodic under $\tau\sim \tau + \beta$.}
\begin{equation}
  \partial_\tau a(\tau)=0\ .
  \label{eq:afix}
\end{equation}
Then the partition function for (\ref{eq:Lfree}) can be written as
\begin{align}
  Z&=\int \mathcal{D}a\mathcal{D}\tilde{A}_0 \mathcal{D}A_i\mathcal{D}\phi 
  \mathcal{D}\lambda\, \delta(\partial_ta)\, \d (\nabla_i A^i)\,\Delta_1[a]\,
  \D_2[A_i]\, \exp(-S[\tilde{A}_0,A_i,\phi,\lambda,a]),\no
  &= \int da_0\mathcal{D}\tilde{A}_0 \mathcal{D}A_i \mathcal{D}\phi 
  \mathcal{D}\lambda\, \d (\nabla_i A^i)\,
  \Delta_1[a_0]\, \Delta_2[A_i]\,\exp(-S[\tilde{A}_0,A_i,\phi,\lambda,a_0])\ ,
\end{align}
where $a_0$ is the zero mode of $a(t)$, which is defined by 
$a_0=\beta^{-1}\int^\beta_0 dt \,a(t)$. 
The Faddeev-Popov determinant can be written as 
\begin{align}
\Delta_1[a] = \text{Det}'(\p_0D_0 ), \qquad \D_2 =
\text{Det}(\nabla^2),
\end{align}
where the domain of the functional determinant $\Delta_1$ is the zero mode of $S^3$ and
the nonzero modes of $S^1$, which is the time direction. 
Because the zero mode of $S^3$ has eigenvalues $m_L=0$ and $m_R=0$
and the gauge field does not have an R-charge, we
can substitute $D_0'=D_0$ in the expression for $\Delta_1[a]$. 
Hence, $\Delta_1[a]$ can be calculated as
\begin{equation}
  \begin{split}
    \Delta_1[a]&= \text{Det}'(\p_0)\cdot \text{Det}'(\partial_0+i[a_0,\ \cdot\ ])=
    \prod_{m\neq 0}\f{2\pi im}{\b}\cdot\prod_{n\neq 0}\prod_{i,j}\left[ \frac{2\pi in}{\beta }+i(\alpha_{i}-
      \alpha_{j})\right]\\
    &=\left( \prod_{m\neq 0}\frac{2\pi im}{\beta }\right)^{N^2+1}
    \prod_{i,j}\prod_{n=1}^\infty\left[1-\frac{\beta^2(\alpha_i-\alpha_j)^2}
      {4\pi^2n^2}\right]\\
    &=\left( \prod_{m\neq 0}\frac{2\pi im}{\beta }\right)^{N^2+1} \prod_{i<j}\frac{4}{\beta ^{2}
      (\alpha_{i}-\alpha_{j})^{2}}\sin ^{2}\left( \frac{\beta (\alpha_{i}-\alpha_{j})}
      {2}\right) ,
  \end{split}
\end{equation}
where $\alpha_i\ (i=1,\dots,N)$ are eigenvalues of $a_0$. 
In the last equality, we have used the infinite product formula
\begin{equation}
  \prod_{n=1}^{\infty}\left(1-\frac{x^2}{n^2}\right)
  =\frac{1}{\pi x}\sin(\pi x)\ .
  \label{eq:infprod}
\end{equation}
The left-right invariant integration measure over Hermitian
matrices $a_0$ is
\begin{equation}
  da_0=\prod_{i}d\alpha_{i}\prod_{i<j}(\alpha_{i}-\alpha_{j})^{2}[d\Om]\ ,
\end{equation}
where $[d\Om]$ is the gauge volume arising from the diagonalization of
$a_0$.
Neglecting this volume, we obtain
\begin{equation}
  da_0\, \Delta_1[a_0] = \prod_id\a_i\, \left[\left( \prod_{m\neq 0}\frac{2\pi im}{\beta }\right)^{N^2+1} 
    \left(\prod_{i<j}\frac{1}{\beta ^{2}}\right)\right]
  \prod_{i<j}4\sin ^{2}\left( \frac{\beta (\alpha_{i}-\alpha_{j})}
    {2}\right)\ .
  \label{eq:daDel}
\end{equation}
The contents of the square bracket in (\ref{eq:daDel}) do not depend on $\alpha_i$ and thus
we can neglect this factor. Then, from Appendix 
\ref{Ap:A}, $da_0 \Delta_1[a]$ is equivalent to 
the Haar measure of $U(N)$. Hence, we denote $da_0\Delta_1[a]$ as $dU$. 
Then, the partition function can be written as
\begin{equation}
  \begin{split}
    Z =\int dU
    \int\mathcal{D}\tilde{A}_0\mathcal{D}A_i\,\d (\nabla_i A^i)\, 
    \Delta_2[A_i]\, e^{-S_\text{gauge}[\tilde{A}_0,A_i,a_0]}
    \int\mathcal{D}\phi\,e^{-S_\text{scalar}[\phi,a_0]}
    \int\mathcal{D}\lambda\,e^{-S_\text{fermion}[\lambda,a_0]}\ ,
    \end{split}
\end{equation}
where $S_\text{gauge}$, $S_\text{scalar}$ and $S_\text{fermion}$ are
the gauge, scalar and spinor field sectors of action~(\ref{eq:Lfree}),
respectively.

First, we focus on the gauge field in the Lagrangian~(\ref{eq:Lfree}).
We can write it as follows after integration by parts:
\begin{align}
  S_\text{gauge} = \int d^4x \s g\ \text{tr}\Big[ -\ti A_0 \nabla^2 \ti A_0 -& 
    A^i \big( (D_0'^2 + \nabla^2)g_{ij} - \CR_{ij}\big) A^j \no
  &+ 2\ti A_0 D_0'\nabla_iA^i + A^j\nabla_j\nabla_i A^i\Big].
\end{align}
The last two terms vanish because of the Coulomb gauge~(\ref{Coulomb}).
The path integral for the gauge field becomes
\begin{align}\label{pathgauge}
  &\int \CD \ti A_0 \CD A_i\, \d (\nabla_i
  A^i)\, \D_2[A_i]\, e^{-S_\text{gauge}[A,a_0]} \no
  &= \int \CD \ti A_0 \CD A_i\, \d (\nabla_i A^i)\, \D_2[A_i]\, \exp \left( 
    \int d^4x \s g\ \text{tr} \left[ \ti A_0 \nabla^2 \ti A_0 +
      A^i \big( D_0'^2 + \nabla^2  - 2\big) A_i \right]
  \right)\no
  &= \text{Det}(\nabla^2)\cdot \text{Det}(\nabla^2)^{-1/2}\int \CD A_i\,  
  \d (\nabla_i A^i)\, \exp \left( \int d^4x \s g\ \text{tr} \left[
      A^i \big( D_0'^2 + \nabla^2  - 2\big) A_i \right]\right),
\end{align}
where we use $\CR_{ij} = \CR g_{ij}/3 = 2g_{ij}$ for $S^3$.
We decompose $A_i$ into a divergenceless vector and a scalar part as
\begin{equation}
 A_i=B_i + \partial_i \varphi,
\end{equation}
where $\nabla_i B^i=0$. By this field redefinition, the measure is
replaced by
$\mathcal{D}A_i=\mathcal{D}B_i\mathcal{D}\varphi\text{Det}(\nabla^2)^{1/2}$,
and Eq.~(\ref{pathgauge}) becomes
\begin{equation}
\begin{split}
&\text{Det}(\nabla^2)\int \CD B_i \CD \varphi\,  
  \d (\nabla^2\varphi)\, \exp \left( \int d^4x \s g\ \text{tr} \left[
      B^i \big( D_0'^2 + \nabla^2  - 2\big) B_i \right]\right)\\
&=\int \CD B_i \,  
  \exp \left( \int d^4x \s g\ \text{tr} \left[
      B^i \big( D_0'^2 + \nabla^2  - 2\big) B_i \right]\right)\\
&=\text{Det}(D_0'^2 + \nabla^2  - 2)^{-1/2}\ ,
\end{split}
\label{eq:gpathint}
\end{equation}
where the functional determinant $\text{Det}(\nabla^2)$ has been canceled
completely. 
In the final expression of (\ref{eq:gpathint}), 
the domain of the functional determinant is a divergenceless vector. 
The eigenvalues of $D_0'$ and $\nabla^2$ for the gauge field are
given by
\begin{equation}
\begin{split}
 &D_0' = \frac{2\pi in}{\beta} + i\alpha_{ij} - \Omega_+ m_L -
  \Omega_- m_R \ ,\\
 &-\nabla^2+2= E_v^2 = (2j+2)^2\ ,
\end{split}
\end{equation}
where $\alpha_{ij}\equiv \alpha_i-\alpha_j$ and $E_v$ is the energy
of the divergenceless vector derived in section \ref{subsec:Spec}.
Since the gauge field does not have an R-charge, we can set
$\hat{Q}_a=0$ in (\ref{eq:D0'}). 
From section \ref{subsec:Spec}, $m_L$ and $m_R$
satisfy $|m_L|\leq j,|m_R|\leq j+1$ for the $(j,j+1)$ representation
or $|m_L|\leq j+1,|m_R|\leq j$ for the $(j+1,j)$ representation. 
Hence, the functional determinant can be calculated as
\begin{equation}
\begin{split}
 &\ln\text{Det}(-D_0'^2-\nabla^2+2)^{-1/2}
=-\f{1}{2}\,\text{Tr}\ln(-D_0'^2-\nabla^2+2)\\
&=-\f{1}{2}\sum_{(j,m_L,m_R)}\sum_{i,j}\sum_{n=-\infty}^\infty
\ln\left[\left(\frac{2\pi n}{\beta} + \alpha_{ij} +i \Omega_+ m_L +i
  \Omega_- m_R \right)^2+E_v^2\right]\\
&=-\f{1}{2}\sum_{(j,m_L,m_R)}\sum_{i,j}\bigg\{
\sum_{n\neq 0}\left[\ln
\left(1+\frac{\beta(\alpha + iE_v)}{2\pi n}\right)
+\ln\left(1+\frac{\beta(\alpha - iE_v)}{2\pi n}\right)
\right]\\
&\hspace*{3.5cm}+\ln\left(\beta^2(\alpha + iE_v)(\a - iE_v)\right)+\sum_{n\neq 0}\ln\left(\frac{2\pi n}{\beta}\right)^2-\ln(\beta^2)
\bigg\},
\end{split}
\label{eq:VDet}
\end{equation}
where we have defined 
$\alpha \equiv \alpha_{ij} +i \Omega_+ m_L +i \Omega_- m_R$.
Because the last two terms in (\ref{eq:VDet}) are constant and 
independent of $\a$ and $E_v$, we will neglect them.
Then, the curly brackets in (\ref{eq:VDet}) become
\begin{align}
      \sum_{n= 1}^\infty&\left[\ln 
        \left( 1-\frac{\beta^2(\alpha + iE_v)^2}{4\pi^2 n^2}\right)
        +\ln\left(1-\frac{\beta^2(\alpha - iE_v)^2}{4\pi^2 n^2}\right)
      \right]
      +\ln\left(\beta^2(\alpha + iE_v)(\a - iE_v)\right)\no
    &= \ln\left[4\sin\left(\frac{\beta(\alpha + iE_v)}{2}\right)
      \sin\left(\frac{\beta(\alpha - iE_v)}{2}\right)\right]\no
    &= \ln\left[e^{\beta E_v}(1-e^{-\b E_v +
        i\beta\alpha})(1-e^{-\b E_v
        -i\beta\alpha})\right]   \label{eq:VDet2}  \notag \\
    &= \beta E_v
    + \ln\left[(1-e^{-\b E_v + i\beta\alpha})(1-e^{-\b E_v -i\beta\alpha})\right]\ .
\end{align}
In the first equality, we have used (\ref{eq:infprod}). 
The first term in the final line of (\ref{eq:VDet2}) represents the Casimir
energy. However, we should not take this term into account 
since, in gravity theory,
we have measured the thermodynamical quantities relative to AdS spacetime, 
and the contribution from the Casimir energy has been subtracted.
Then, (\ref{eq:VDet}) becomes
\begin{align}
    -\f12\sum_{(j,m_L,m_R)}&\sum_{i,j} \left[\ln(1-e^{-\b E_v + i\beta\alpha}) +
    \ln( 1-e^{-\b E_v - i\beta\alpha})\right] \no
    &=\f12\sum_{(j,m_L,m_R)}\sum_{i,j}\sum_{n=1}^\infty 
    \frac{1}{n}e^{-n\beta (E_v - i\alpha)} + (\a \to -\a) \no
    &=\f12\sum_{(j,m_L,m_R)}\sum_{n=1}^\infty\frac{1}{n}
    e^{-n\beta(E_v+\Omega_+ m_L + \Omega_- m_R )}
    \sum_{i=1}^Ne^{in\beta\alpha_{i}}\sum_{j=1}^Ne^{-in\beta\alpha_{j}} + (\a \to -\a)\no
    &=\sum_{n=1}^\infty\frac{z_V(x^n,\mu_a,\Omega_i)}{n} \text{tr}(U^n)
    \text{tr}(U^{-n}) ,
\end{align}
where $x=e^{-\beta}$ and $U=e^{i\beta a_0}$.
$z_V(x,\mu_a,\Omega_i)$ is the single-particle partition function
of the gauge field defined by (\ref{PFVCR}).
Therefore, the path integral for the gauge field is given by
\begin{equation}
  \int\mathcal{D}\tilde{A}_0\mathcal{D}A_i\,\d (\nabla_i A^i)\, 
  \Delta_2[A_i]\, e^{-S_\text{gauge}[\tilde{A}_0,A_i,a_0]}
  =\exp\left(\sum_{n=1}^\infty\frac{z_V(x^n,\mu_a,\Omega_i)}{n} 
    \text{tr}(U^n) \text{tr}(U^{-n})\right).
\end{equation}

Next, we consider the scalar fields in the action~(\ref{eq:Lfree}). 
The path integral for scalar fields is
\begin{equation}
\int\mathcal{D}\phi\,e^{-S_\text{scalar}[\phi,a_0]}
=\text{Det}(-D_0'^2-\nabla^2+1)^{-1/2}\ .
\end{equation}
The eigenvalues of $D'_0$ and $\nabla^2$ for the scalar fields are given by
\begin{align}
 &D_0'=\frac{2\pi in}{\beta}+i\alpha_{ij}-\sum_{p=1}^3\mu_a Q_a
 - \Omega_+ m_L - \Omega_- m_R\ ,\notag \\
 &-\nabla^2+1= E_s^2=(2j+1)^2\ ,
\end{align}
where $Q_a=\pm 1$ and $|m_L|,|m_R|\leq j$.
By a similar calculation to that in the gauge field case, the
path integral for the scalar fields can be written as
\begin{equation}
 \int\mathcal{D}\phi\,e^{-S_\text{scalar}[\phi,a_0]}
=\exp\left(\sum_{n=1}^\infty\frac{z_S(x^n,\mu_a,\Omega_i)}{n}
  \text{tr}(U^n) \text{tr}(U^{-n})\right) ,
\end{equation}
where $z_S(x,\mu_a,\Omega_i)$ is the single-particle partition function 
for the scalar field defined by (\ref{PFSCR}).

Finally, we consider the fermions. 
The path integral for fermions is
\begin{align}
 &\int\mathcal{D}\lambda\,e^{-S_\text{fermion}[\lambda,a_0]}
= \text{Det}(i\Dslash')=\text{Det}(-\Dslash'^2)^{1/2}\ , \notag \\
&=\text{Det}\left(-D_0'^2-\nabla^2+\frac{3}{2}\right)^{1/2}.
\end{align}
The eigenvalues for $D'_0$ and $\nabla^2$ are given by
\begin{align}
 &D_0'=\frac{(2n+1)\pi i}{\beta}+i\alpha_{ij}-\sum_{p=1}^3\mu_a Q_a
 - \Omega_+ m_L - \Omega_- m_R, \notag \\
 &-\nabla^2+\frac{3}{2}=E_f^2=\left(2j+\frac{3}{2}\right)^2\ ,
\end{align}
where $Q_a=\pm 1/2$.
From section \ref{subsec:Spec}, $m_L$ and $m_R$
satisfy $|m_L|\leq j,|m_R|\leq j+1/2$ for the $(j,j+1/2)$ representation
or $|m_L|\leq j+1/2,|m_R|\leq j$ for the $(j+1/2,j)$ representation. 
Since the fermions are antiperiodic on $S^1$, the
eigenvalue of $\partial_0$ becomes $(2n+1)\pi i/\beta$ and 
the calculation of the functional determinant is slightly different
from that of boson fields. 
\begin{equation}
\begin{split}
 &\ln\text{Det}(-{D'_0}^2-\nabla^2+3/2)^{1/2}
=\frac{1}{2}\, \text{Tr}\ln(-{D'_0}^2-\nabla^2+3/2)\\
&=\frac{1}{2}\sum_{(j,m_L,m_R,Q_a)}\sum_{i,j}\bigg\{
\sum_{n=-\infty}^\infty\left[\ln
\left(1+\frac{\beta(\alpha + iE_f)}{2\pi (n+1/2)}\right)
+\ln\left(1+\frac{\beta(\alpha - iE_f)}{2\pi (n+1/2)}\right)
\right]
+\ln(4)\\
&\hspace*{8.5cm}
+\sum_{n=-\infty}^\infty\ln\left(\frac{2\pi (n+1/2)}{\beta}\right)^2-\ln(4)
\bigg\}.
\end{split}
\label{eq:MDet}
\end{equation}
Because the last two terms in (\ref{eq:MDet}) are constant, 
we will neglect them.
Then, the curly brackets in the above equation become
\begin{align}
&\sum_{n= 1}^\infty\left[\ln
\left(1-\frac{\beta^2(\alpha + iE_f)^2}{4\pi^2 (n-1/2)^2}\right)
+\ln\left(1-\frac{\beta^2(\alpha - iE_f)^2}{4\pi^2 (n-1/2)^2}\right)
\right]
+\ln(4)\no
&=
 \ln\left[4\cos\left(\frac{\beta(\alpha + iE_f)}{2}\right)
\cos\left(\frac{\beta(\alpha - iE_f)}{2}\right)\right]\no
&=
 \ln\left[e^{\beta E_f}(1+e^{-\beta E_f + i\b\alpha
   })(1+e^{-\beta E_f - i\b\alpha})\right]\no
&=\beta E_f
+ \ln\left[(1+e^{-\beta E_f + i\b\alpha})(1+e^{-\beta E_f - i\b\alpha})\right]\ .
\label{eq:MDet2}
\end{align}
In the first equality, we have used the infinite product formula
\begin{equation}
 \prod_{n=1}^\infty\left(1-\frac{x^2}{(n-1/2)^2}\right)=\cos(\pi x)\ .
\end{equation}
The first term in (\ref{eq:MDet2}) represents the Casimir
energy, which we neglect. 
In the second term, we can replace $-i\b\alpha$ by $+i\b\alpha$ in the second
parentheses because of the summations over $(i,j)$  and $(j,m_L,m_R, Q_a)$.
Then, (\ref{eq:MDet}) becomes
\begin{equation}
\sum_{(j,m_L,m_R,Q_a)}\sum_{i,j}\ln(1+e^{-\b E_f +
    i\beta\alpha})
=\sum_{n=1}^\infty\frac{(-1)^{n+1}z_F(x^n,\mu_a,\Omega_i)}{n}\,
\text{tr}(U^n)\,\text{tr}\,(U^{-n})\ ,
\end{equation}
where $z_F(x,\mu_a,\Omega_i)$ is the single-particle partition
function for fermions defined by (\ref{PFFCR}).

Therefore, the partition function taking into account the gauge, scalar and
Majorana fields is given by
\begin{equation}
 Z=\int
  dU\,\exp\left[\sum_{n=1}^\infty\frac{1}{n}\{z_B(x^n,\mu_a,\Omega_i)+(-1)^{n+1}z_F(x^n,\mu_p,\Omega_i)\}
\,\text{tr}(U^n)\,\text{tr}(U^{-n})\right]\ ,
\end{equation}
where $z_B(x,\mu_a,\Omega_i)\equiv
z_V(x,\mu_a,\Omega_i)+z_S(x,\mu_a,\Omega_i)$. 
This expression is equal to (\ref{UMM}), which was derived by a 
group theoretical approach.


\end{document}